\pdfoutput=1
\documentclass[cernpreprint, atlasdraft=false, texlive=2023, UKenglish, texmf, orcidlogo]{atlasdoc}
 
\usepackage{atlaspackage}
\usepackage{atlasbiblatex}
 
\usepackage{atlasphysics}

\graphicspath{{logos/}{figures/}}
 
\usepackage{ANA-HION-2021-16-PAPER-defs}
\usepackage{multirow}

\AtlasTitle{Exclusive dielectron production in ultraperipheral Pb+Pb
collisions at $\ensuremath{\sqrt{s_{_\text{NN}}}}\xspace = 5.02$~TeV with ATLAS}
 
\AtlasVersion{2.2}
 
\AtlasAbstract{
Exclusive production of dielectron pairs, \mbox{$\gamma\gamma\rightarrow e^+e^-$}\xspace, is studied using \mbox{$\mathcal{L}_\mathrm{int}=1.72\;\mathrm{nb^{-1}}$}\xspace
of data from ultraperipheral collisions of lead nuclei at
$\ensuremath{\sqrt{s_{_\text{NN}}}}\xspace = 5.02$~TeV recorded by the ATLAS detector at the LHC. The
process of interest proceeds via photon--photon interactions in the
strong electromagnetic fields of relativistic lead nuclei.
Dielectron production is measured in the fiducial region defined by following
requirements: electron transverse momentum $\ensuremath{p_{\textrm{T}}^e}\xspace > 2.5$~GeV, absolute
electron pseudorapidity $\ensuremath{|\eta^{e}|}\xspace < 2.5$, dielectron invariant mass $\ensuremath{m_{ee}}\xspace > 5$~GeV,
and dielectron transverse momentum $\ensuremath{p_{\textrm{T}}^{ee}}\xspace < 2$~GeV.
Differential cross-sections are measured as a function of \ensuremath{m_{ee}}\xspace,
average \ensuremath{p_{\textrm{T}}^e}\xspace, absolute dielectron rapidity \ensuremath{|y_{ee}|}\xspace, and scattering angle
in the dielectron rest frame, \ensuremath{|\cos\theta^*|}\xspace, in the inclusive sample, and also with a
requirement of no activity in the forward direction.
The total integrated fiducial cross-section is measured to be
$215 \pm 1\text{(stat.)}~\substack{+23\\-20}\text{(syst.)} \pm 4\text{(lumi.)}~\mu$b. Within experimental
uncertainties the measured integrated cross-section is in good agreement
with the QED predictions from the Monte Carlo programs \textsc{Starlight}\xspace and \textsc{SuperChic}\xspace, confirming
the broad features of the initial photon fluxes. The
differential cross-sections show systematic differences from these
predictions which are more pronounced at high \ensuremath{|y_{ee}|}\xspace and \ensuremath{|\cos\theta^*|}\xspace values.
 
}

\AtlasRefCode{HION-2021-16}
 
\PreprintIdNumber{CERN-EP-2022-130}

\AtlasJournal{JHEP}
\AtlasJournalRef{JHEP 06 (2023) 182}
\AtlasDOI{10.1007/JHEP06(2023)182}

\hypersetup{pdftitle={ATLAS document},pdfauthor={The ATLAS Collaboration}}
 
\begin{document}
 
\maketitle
 
\tableofcontents

\section{Introduction}

Collisions of ultrarelativistic heavy ions provide an opportunity to
study not only the strong interactions between nucleons but also
processes involving electromagnetic (EM) interactions. This is due to
the presence of intense EM fields associated with the
ultrarelativistic ions. The EM interactions become dominant at large
impact parameters, $b > 2 R_{A}$, where $R_A$ is the ion radius. Such
collisions are usually referred to as ultraperipheral
collisions~(UPC). Comprehensive reviews of UPC physics can be found
in Refs.~\cite{BALTZ_2008,Klein:2020fmr}.
 
The EM fields associated with the ultrarelativistic nuclei can be
treated as fluxes of quasi-real photons according to the equivalent
photon approximation (EPA) formalism~\cite{Fermi:1925fq,
PhysRev.45.729}. In this approach, the total cross-section for a given
process is calculated as a convolution of the photon flux with the
elementary production cross-section.  Although the same approach is
also valid for proton--proton~($pp$) collisions, the expected cross-sections are
strongly enhanced in the heavy-ion (HI) collisions.  The photon flux
from each nucleus is enhanced by a factor of $Z^{2}$, where $Z$ is the atomic number.
That results in a $Z^4$ enhancement of the cross-sections.
For lead (Pb, $Z=82$), this $Z^4$ enhancement is $4.5\times 10^7$. Another
advantage of studying photon-induced interactions in UPC HI collisions
is the relatively low number of interactions per LHC bunch crossing.
The mean number of simultaneous hadronic interactions, $\mu$, is typically at the
subpercent level. This provides a clean environment, facilitating the
detection of the interaction products, and little contamination from
unrelated interactions in the same crossing.  With the centre-of-mass energy per
nucleon pair available at the Large Hadron Collider~(LHC), $\sqn =
5.02$\,\TeV, the initial photons can reach energies up to 75--100\,\GeV\ in the Pb+Pb centre-of-mass system.
 
\begin{figure}[!t]
\begin{center}
\includegraphics[width=0.3\textwidth]{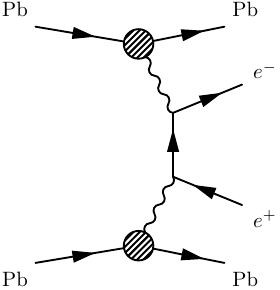}
\caption{Feynman diagram of the leading-order \yyee process.}
\label{fig:FeynmanDiag}
\end{center}
\end{figure}

Among the possible set of photon-induced reactions, the exclusive production
of dielectron pairs from photon--photon collisions, i.e.\
$\gamma\gamma\rightarrow e^+e^-$, is one of the cleanest
elementary processes.
This process, also
referred to as the Breit--Wheeler process~\cite{PhysRev.46.1087}, is a
non-resonant two-photon scattering to opposite-charge electron pairs.
The outgoing nuclei may be excited, due to incoherent emission of photons
participating in \ggee process or via photonuclear reactions, occurring
when an additional photon is exchanged.~\cite{Pshenichnov:2011zz,VEYSSIERE1970561}. A Feynman diagram of the leading-order \yyee reaction is
shown in Figure~\ref{fig:FeynmanDiag}.
Even for large $\gamma\gamma$ invariant masses,
the relatively large cross-section associated with this
process allows precise differential measurements to be made.
Thus, this process is a particularly effective tool for studying the modelling
of photon fluxes and elementary production cross-sections, as well as
for studying the effects of nuclear break-up induced by additional photon
exchanges, whose probability is
strongly correlated with the internuclear impact parameter~\cite{Klein:2020fmr}.
Nuclear break-up gives rise to forward neutron production, and
the fraction of events with such activity is larger at smaller impact parameters.
 
Exclusive dilepton production, via both electron-pair and muon-pair final states,
has been measured by ATLAS and CMS in $pp$ collisions
at $\sqrt{s}=7$\,\TeV~\cite{ATLAS:2015wnx, CMS:2012cve, CMS:2011vma} and
$\sqrt{s}=13$\,\TeV~\cite{ATLAS:2017sfe,ATLAS:2020mve,CMS:2018uvs}. The
ALICE Collaboration has measured exclusive production of electron pairs in
Pb+Pb collisions at
$\sqn=2.76$\,\TeV~\cite{ALICE:2013wjo} over a limited
kinematic range. The STAR and PHENIX experiments at RHIC have measured exclusive
dileptons at lower invariant masses at $\sqn=200$\,\GeV\
for both Au+Au and U+U collisions~\cite{STAR:2018ldd,STAR:2019wlg, PHENIX:2009xtn}. With the
higher centre-of-mass energy of $\sqn=5.02$\,\TeV, ATLAS has
performed differential measurements of
$\gamma\gamma\rightarrow \mu^+\mu^-$ production in UPC Pb+Pb
collisions~\cite{ATLAS:2020epq}. Both STAR and ATLAS have observed
substantial broadening of angular distributions for exclusive
dileptons from $\gamma\gamma$ interactions in events where the
nuclei overlapped and interacted
hadronically~\cite{STAR:2018ldd,ATLAS:2018pfw}.
Finally, CMS has observed that angular
correlations in exclusive dimuon events are broadened significantly
as a function of the impact parameter~\cite{CMS:2020skx},
as inferred by the amount of forward neutron production.
 
Exclusive dielectron production is an important reference process
for other measurements.
References~\cite{DYNDAL2020135682} and~\cite{PhysRevD.102.113008} propose using it in the context of \ggtautau production,
in order to reduce the impact of correlated systematic uncertainties for the
measurement of the $\tau$-lepton anomalous magnetic moment.
It is also an important background for light-by-light scattering, which
proceeds via loop diagrams and thus has a much lower cross-section.
This has been assessed in several publications on light-by-light scattering
by ATLAS~\cite{ATLAS:2017fur,ATLAS:2019azn,ATLAS:2020hii} and CMS~\cite{CMS:2018erd}.

This paper presents a measurement of the exclusive production of dielectrons with
the ATLAS detector at the LHC. It uses \PbPb
data collected in 2018, which have an integrated luminosity three times
larger than the sample used in the ATLAS dimuon
measurement~\cite{ATLAS:2020epq}.
Dielectron production is measured in the fiducial region defined by the following
requirements: electron transverse momentum $\pTe > 2.5$\,\GeV, electron
pseudorapidity  $\etae < 2.5$, dielectron invariant mass $\mee > 5$\,\GeV,
and dielectron transverse momentum $\pTee < 2$\,\GeV. Compared to the dimuon measurement discussed in
Ref.~\cite{ATLAS:2020epq}, this fiducial region has wider coverage
in lepton \pt and dilepton invariant mass, with the minimum values
lowered by 1.5\,\GeV\ and 5\,\GeV, respectively.
The backgrounds originating from single-dissociative processes,
$\Upsilon(nS)$ production, and exclusive $\tau$-lepton pair production,
$\gamma\gamma\rightarrow\tau^+\tau^-$, are estimated and subtracted.
Differential cross-sections are measured as a function of \mee, average
electron transverse momentum \apTe, absolute dielectron rapidity \Yee, and
scattering angle in the dielectron rest frame, \costs. The cross-sections are
extracted both inclusively in forward neutron activity and exclusively in \ggee events
without activity in the forward direction. The latter is a unique feature of this paper.
 
\section{ATLAS detector}

\newcommand{\AtlasCoordFootnote}{
ATLAS uses a right-handed coordinate system with its origin at the nominal interaction point (IP)
in the centre of the detector and the $z$-axis along the beam pipe.
The $x$-axis points from the IP to the centre of the LHC ring,
and the $y$-axis points upwards.
Cylindrical coordinates $(r,\phi)$ are used in the transverse plane,
$\phi$ being the azimuthal angle around the $z$-axis.
The pseudorapidity is defined in terms of the polar angle $\theta$ as $\eta = -\ln \tan(\theta/2)$.
Angular distance is measured in units of $\Delta R \equiv \sqrt{(\Delta\eta)^{2} + (\Delta\phi)^{2}}$.
The photon (electron) transverse energy is $\et= E/\cosh(\eta)$, where $E$ is its energy.}
 
The ATLAS detector~\cite{ATLAS_detector_paper} at the LHC covers nearly the entire solid angle around the collision point.
It consists of an inner tracking detector surrounded by a thin superconducting solenoid, electromagnetic and hadronic calorimeters,
and a muon spectrometer incorporating three large superconducting
toroidal magnets with eight coils each.
 
The inner detector (ID) is immersed in a \SI{2}{\tesla} axial magnetic field
and provides charged-particle tracking in the pseudorapidity range $|\eta| < 2.5$.\footnote{\AtlasCoordFootnote}
The high-granularity silicon pixel~(Pixel) detector covers the vertex region and typically provides four measurements per track,
with the first hit normally being in the insertable B-layer~(IBL)~\cite{ibl_Run2}, which was installed at the mean distance of 3.3\,cm from the beam axis before the start of Run~2.
It is followed by the silicon microstrip tracker~(SCT), which usually provides four two-dimensional measurement points per track.
These silicon detectors are complemented by the transition radiation tracker,
which enables radially extended track reconstruction up to $|\eta| = 2.0$.
 
The calorimeter system covers the pseudorapidity range $|\eta| < 4.9$.
Within the region $|\eta|< 3.2$, electromagnetic calorimetry is provided by barrel and
endcap lead/liquid-argon~(LAr) calorimeters (high granularity for $|\eta| < 2.5$),
with an additional thin LAr presampler covering $|\eta| < 1.8$
to correct for energy loss in material upstream of the calorimeters.
Hadronic calorimetry is provided by the steel/scintillator-tile calorimeter, which is
segmented into three barrel structures within $|\eta| < 1.7$, and two copper/LAr hadronic endcap calorimeters.
The solid angle coverage is completed with forward copper/LAr and
tungsten/LAr calorimeter modules~(FCal) optimised for electromagnetic and hadronic measurements respectively.
 
The muon spectrometer~(MS) comprises separate trigger and
high-precision tracking chambers measuring the deflection of muons in
a magnetic field generated by the superconducting air-core toroids.  The
precision chamber system covers the region $|\eta| < 2.7$ with three
layers of monitored drift tubes, complemented by cathode strip
chambers in the forward region, where the background is highest.  The
muon trigger system covers the range $|\eta| < 2.4$ with resistive
plate chambers in the barrel, and thin gap chambers in the endcap
regions.
 
The zero-degree calorimeters~(ZDC) consist of four longitudinal
compartments on each side of the IP, each with one nuclear interaction
length of tungsten absorber, with Cherenkov light read out by
1.5-mm-diameter quartz rods.  The detectors are located 140\,m from the nominal
IP in both directions, covering $|\eta|>8.3$, and are well suited to
measuring neutral particles originating from the collision.  In \PbPb
collisions the ZDC detects individual neutrons originating
from the incoming nuclei. The ZDC calibration is performed in each
set of four modules using photonuclear processes that deposit one
or more neutrons on one side, and a single neutron,
carrying the full per-nucleon beam energy, on the other. Time-dependent
weights are determined for each module in short time intervals to minimise the
variance around the nominal per-nucleon beam energy. Energy resolutions
achieved are typically around $\Delta E / E \approx$ 16\%.
 
The ATLAS trigger system~\cite{ATLAS:2016wtr,TRIG-2019-04} consists of a first-level (L1)
trigger implemented using a combination of dedicated electronics and
programmable logic, and a software-based high-level trigger~(HLT).
An extensive software suite~\cite{ATL-SOFT-PUB-2021-001} is used in
the reconstruction and analysis of real and simulated data, in
detector operations, and in the trigger and data acquisition systems
of the experiment.
 
\section{Data and Monte Carlo simulation samples}
 
The data used in this measurement are from \PbPb collisions with a
centre-of-mass energy of $\sqn = 5.02$\,\TeV, recorded in 2018 with the
ATLAS detector at the LHC. The full data set corresponds to an
integrated luminosity of 1.72\,nb$^{-1}$. Only high-quality data~\cite{DAPR-2018-01} with
all detectors operating normally are analysed.
 
Monte Carlo~(MC) simulated events for the \ggee signal process were
generated at leading order~(LO) using
\starlight~v3.13~\cite{Klein:2016yzr}. In this approach, the cross-section is
computed by convolving the Pb+Pb photon flux with the LO calculation
of the elementary \ggee process.  The photon spectrum is calculated
in impact parameter space by integrating the photon number density
over all impact parameters while assuming that the beam
projectiles do not interact hadronically.  This is done by utilising a simple
Glauber model~\cite{doi:10.1146/annurev.nucl.57.090506.123020}
which provides an impact-parameter-dependent probability
of inelastic processes.
\starlight also requires that the dilepton pairs are not formed within
either nucleus.
Several signal samples were produced for exclusive \mee intervals
within the range $4.5< m_{ee} < 200$\,\GeV. An alternative sample for the signal
\ggee\ process uses the \superchic~v3.05~\cite{Superchic3} program. The difference
between the nominal and alternative signal prediction is mainly in the
implementation of the initial photon flux. The \starlight generator
relies on flux from point-like sources restricted to impact parameters
larger than nuclear radius ($b>R_A$), while \superchic generator implements
flux calculations down to $b=0$ and takes the nuclear form factor into account.

Backgrounds from \ggtautau\ and $\Upsilon(nS)\rightarrow e^+ e^-$
were simulated using \starlight~v3.13.
All generated events utilised \pythia~(Py8)~\cite{Sjostrand:2014zea} to
model QED final-state radiation~(FSR) from the outgoing leptons.
 
Incoherent emission of photons participating in \ggee process
contribute to background in the dielectron production sample.
In such a process either one (single dissociation) or both
(double dissociation) nuclei interact inelastically and dissociate.
Simulation of this process is not available in any MC generator, however
a similar process in \pp collisions is modelled using
\superchic~v4.0~(SC4)~\cite{Harland-Lang:2020veo}.
The sample generated with SC4 was interfaced to \pythia for showering
and hadronisation. A data-driven approach, discussed in detail in
Section~\ref{sec:bkg}, is used in the analysis to
utilise this sample in \PbPb collisions.
 
Apart from the alternative signal sample, all generated events are processed with a
detector simulation~\cite{Aad:2010ah} based on \GEANT~\cite{Agostinelli:2002hh} and are reconstructed with
the standard ATLAS reconstruction software~\cite{ATL-SOFT-PUB-2021-001}.
The alternative signal sample is used for comparisons with the
measured differential cross-sections discussed in Section~\ref{sec:result}.
 
\section{Signal selection and detector corrections}
\label{sec:analysis}
Candidate dielectron events were recorded using a dedicated trigger
for events with moderate activity in the calorimeter but little
additional activity in the entire detector. A logical OR of two L1 trigger conditions
was required: (1) at least one EM cluster with $\et > 1$\,\GeV\ in
coincidence with a total \et of 4--200\,\GeV\ registered in the calorimeter,
or (2) at least two EM clusters with $\et > 1$\,\GeV\ and a total \et below 50\,\GeV\ registered
in the calorimeter. At the HLT, the total \et on each side of the FCal
detector was required to be below 3\,\GeV.
Additionally, a veto condition on the maximum activity in the Pixel
detector, hereafter referred to as the Pixel-veto, had to be
satisfied at the HLT. The number of hits was required to be at most 15 to be
compatible with low-multiplicity UPC events.
 
Electrons are reconstructed from EM clusters in the calorimeter and
tracking information provided by the ID~\cite{ATLAS:2019qmc}.
Selection requirements are applied to remove EM clusters with a large
amount of energy from poorly functioning calorimeter cells, and a
timing requirement is made to reject out-of-time candidates. An energy
calibration specifically optimised for electrons and
photons~\cite{ATLAS:2019qmc} is applied to the candidates to account
for upstream energy loss and both lateral and longitudinal shower
leakage. The calibration is derived for nominal \pp collisions with
dedicated factors applied to account for the much lower contribution
from multiple \PbPb collisions in the same bunch crossing.
 
The electron identification in this analysis is based on a `loose'
cut-based working point~\cite{ATLAS:2019qmc} which is defined using
selections on the shower-shape and tracking variables. Only electrons
with $\pTe > 2.5$\,\GeV\ and $\etae < 2.47$, excluding the
calorimeter transition region $1.37 < \etae < 1.52$, are considered.
The minimum \pTe requirement is driven by the electron reconstruction
efficiency, which drops below 20\% for \pTe values below this threshold.
 
Preselected events are required to have exactly two opposite-charge
electrons satisfying the above selection criteria, with a dielectron
invariant mass, \mee, greater than 5\,\GeV. To suppress non-exclusive
backgrounds, only two charged-particle tracks~\cite{ATL-PHYS-PUB-2015-051,PERF-2015-08} each
with $\pt > 100$\,\MeV, $|\eta| < 2.5$, at least seven hits in the Pixel and SCT detectors in
total and at most two silicon sensors without a hit, and associated
with the dielectron are allowed. To reject
non-collision backgrounds such as cosmic-ray muons, the event must not
have a track in the MS. Finally, the total \pt of
the dielectron, \pTee, is required to be less than 2\,\GeV. Low
\pTee values are a key feature of the purely EM process, which
involves initial-state photons with very low \pT.

\begin{figure}[!htb]
\begin{center}
\includegraphics[width=0.62\textwidth]{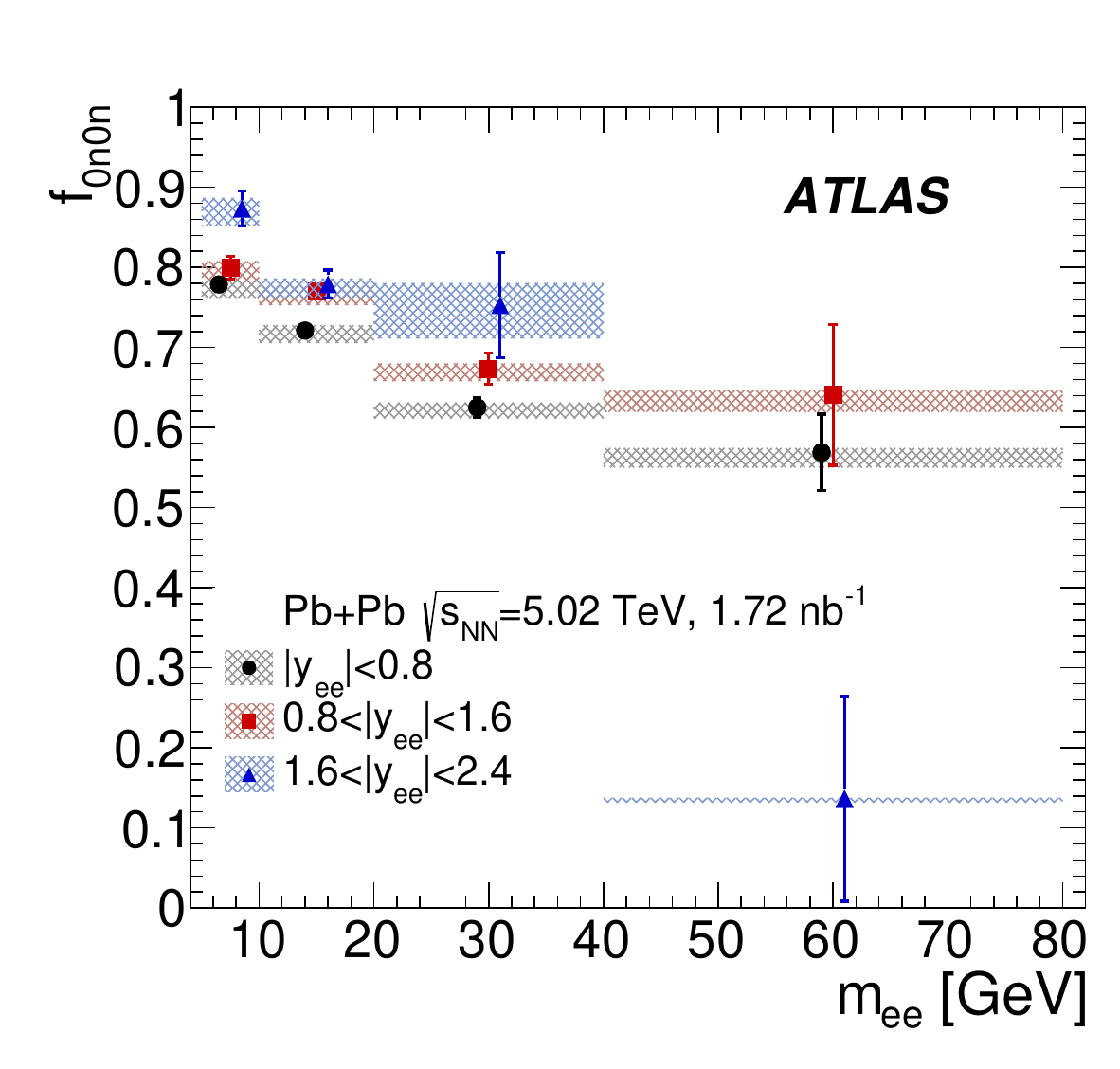}
\caption{Fractions of events in the 0n0n category evaluated from data in
three bins of \Yee, corrected for the
presence of additional neutrons.
Error bars represent statistical uncertainties, while shaded boxes represent
systematic uncertainties. Points for $\Yee<0.8$ and
$1.6<\Yee<2.4$ are displaced horizontally for better
visibility.}
\label{fig:f0n0n}
\end{center}
\end{figure}

Each of the events satisfying the \ggee criteria can be
further classified into one of three categories based on the observed activity in the ZDC detector:
1) no neutron is registered in either ZDC (`0n0n'),
2) one or more forward neutrons registered in one ZDC and none in the other
(`Xn0n'),
and 3) one or more forward neutrons detected in both ZDC arms (`XnXn').
The observed fractions of events falling into these categories are:
$f_\mathrm{0n0n}=(62.9\pm0.3)$\%, $f_\mathrm{Xn0n}=(29.7\pm0.3)$\%, and $f_\mathrm{XnXn}=(7.4\pm0.2)$\%.
Due to the relatively large instantaneous luminosity of \PbPb collisions,
which peaked around $7\times 10^{27}$\,cm$^{-2}$s$^{-1}$,
additional neutrons might be generated per bunch crossing
by single and mutual dissociation processes
and detected in one or both arms of the ZDC,
but they are not associated with the exclusive dielectron process.
This leads to an outflow of events from the 0n0n and Xn0n categories to both the Xn0n and XnXn
categories. This effect is accounted for using the method established in
Ref.~\cite{ATLAS:2020epq}. A matrix equation with two fundamental
parameters representing probabilities for single and mutual dissociation is built.
The corrected fractions are measured in four bins of \mee, with boundaries at 5, 10, 20, and 40\,\GeV, and three bins of \Yee, with
boundaries at 0, 0.8, 1.6, and 2.4, and also in the sample integrated
over $m_{ee}$ and $|y_{ee}|$. On average, in the 0n0n category, they are about 13\% larger than the
observed fractions.
Figure~\ref{fig:f0n0n} shows the fractions of events in the 0n0n
category as a function of \mee in three bins of \Yee,
corrected for the presence of additional neutrons. These fractions
tend to drop with increasing mass, and are in general larger for higher \Yee values.
For the rapidity range of $\Yee<0.8$,
which has the largest number of events, the $f_\mathrm{0n0n}$ values drop from about 78\%
in the lowest mass bin to about 57\% in the highest mass bin. The
systematic uncertainties in the fractions of events in the 0n0n
category originate from several sources: uncertainties in the
exclusive single and double EM dissociation cross-sections measured by the ALICE
Collaboration~\cite{ALICE:2012aa}, and their extrapolation from $\sqn = 2.76$\,\TeV\
to 5.02\,\TeV\ as evaluated in Ref.~\cite{ATLAS:2020epq}; the uncertainty
in the dissociative background contribution as discussed in
Section~\ref{sec:bkg}; and the uncertainty in the ZDC efficiency.
 
The efficiency of the primary physics trigger ($\epsilon_\mathrm{T}$) is
determined as $\epsilon_\mathrm{T} = \varepsilon_{\mathrm{L1}} \cdot
\varepsilon_{\mathrm{PixVeto}} \cdot  \varepsilon_{\mathrm{FCal}}$, where
$\varepsilon_{\mathrm{L1}}$ is the efficiency of the L1 EM
trigger to register the moderate calorimeter activity characteristic
of the signal process, $\varepsilon_{\mathrm{PixVeto}}$ is the efficiency of the
trigger to reject events with large numbers of Pixel detector hits,
and $\varepsilon_{\mathrm{FCal}}$ is the efficiency of the FCal
selection to reject events with large energy depositions on either
side.  Individual efficiencies are evaluated in a sample of \ggee
events collected with a set of dedicated supporting triggers
that do not use the condition under study in the primary physics trigger
to reject any events. The $\varepsilon_{\mathrm{L1}}$ value rises with
the sum of the transverse energies of the two electron clusters and
reaches 100\% for $\Sigma \ET > 8$\,\GeV.  The Pixel-veto efficiency
and its uncertainties are measured as a function of the dielectron rapidity in the dedicated
dielectron sample collected by the supporting trigger requiring at
least two tracks with \pt above 1\,\GeV\ and without veto requirement
on the number of Pixel hits. The efficiency is evaluated as a ratio of
events passing the Pixel-veto requirement to all selected events. It is just over
80\% for $\Yee \sim 0$ and falls to about 50\% for $\Yee> 2$. The
dependence on $\Yee$ originates from the growing number of
Pixel-detector layers in the forward direction that a dielectron pair has to pass through.
The average number of Pixel hits for the signal process is 10 in the central region,
and it increases to 14 in the forward direction. The systematic uncertainties are evaluated
by repeating the efficiency measurement on the dielectron sample selected by varying
requirements on \pTe and \mee, and acoplanarity ($\alpha=1-|\Delta\phi|/\pi$,
where $\Delta \phi$ is the azimuthal angle between the two electrons).
They are at the level of 0.5\%, while the statistical uncertainty ranges between 1 and 2\%.
Finally, the FCal veto efficiency is measured to be $(99.1\pm 0.6)\%$,
and it is constant for the entire range in $\Sigma \ET$. The total
uncertainty in the trigger efficiency is determined by increasing and decreasing all of
the individual components by their respective total
uncertainties. They amount to about 3\%--4\% for the primary calorimeter pair
trigger, driven mainly by the limited number of dielectron events collected by an
independent ZDC-based trigger used to measure $\varepsilon_{\mathrm{L1}}$,
and less than a percent for the other contributions.
 
The total electron efficiency is the product of the electron reconstruction efficiency and the
`loose' electron identification efficiency~\cite{ATLAS:2019qmc}. This is determined in data using a sample of events triggered
by the presence of EM clusters, limited total $E_{\mathrm{T}}$, and a maximal number of Pixel hits,
on which a tag-and-probe procedure is performed.  The tag is a well-reconstructed, high-purity electron candidate
with $\ET>2.5$\,\GeV, and the probe is an opposite-charge track built from at least three hits in the
Pixel detector (referred to as a `Pixel-track'). Pixel tracks are required to have $\pt > 50$\,MeV. The invariant mass of the tag-and-probe system must
exceed 5\,\GeV\ and the acoplanarity has to be less than 0.1.
The extracted mass distribution is found to agree well with a
reconstructed sample of \starlight\ events.
The reconstruction efficiency is defined as the fraction of probes which
are reconstructed electrons, while the identification efficiency is the
fraction of reconstructed electrons which are identified as `loose'
electrons.
The reconstruction efficiency has large variations with both $\pT$ and
Pixel-track $\eta$,
and ranges from about 30\% at $\pT=2.5$\,\GeV\ to 95\% above 15\,\GeV.
The identification efficiency is found to vary more weakly with
Pixel-track $\eta$, ranging between 80\% and 90\%.
Then, the overall reconstruction scale factors are extracted as the ratio of efficiencies
measured in data and MC simulation. They vary between 0.9 and 1.2, with the largest
deviations from unity being in the forward direction for
Pixel-track $|\eta| > 1.1$. Systematic uncertainties in the scale
factors are evaluated using tighter selection criteria for the tag
and probe candidates, as well as reducing a potential contribution
from background processes by limiting the measurement to the 0n0n category
or to a narrow acoplanarity region, $\alpha<0.01$. In particular the requirement on
maximum $\alpha$ reduces the contribution from the exclusive $\tau^{+}\tau^{-}$
production to 0.25\% level. The total systematic
uncertainty is at the level of 5\% for central Pixel-tracks with $|\eta|<1$, and
grows to 10\% in the forward direction. In the forward region, the
statistical and systematic uncertainties are of similar size.

\section{Background contributions}
\label{sec:bkg}
There are three primary sources of background considered in this analysis,
presented in order of decreasing contributions: dissociative \ggee production;
$\Upsilon$-meson production; and exclusive $\tau$-lepton pair ($\tau^+\tau^-$) production.

The largest background originates from \ggee production with
nuclear dissociation. In this process one (or both) of the initial photons
originates from the substructure of the nucleon, rather than from
the exterior EM field of the nuclei as a whole. The photon interaction that produces
the $e^+e^-$ pair is thus accompanied by
the dissociation of the emitting nucleus, whose remnants are produced in the
forward direction and are typically captured by the ZDC detector.
 
The contribution from dissociative events is estimated using a
template-fitting approach applied to the acoplanarity
distribution. The signal template is
simulated with \starlightPythia and it is peaked at $\alpha \approx 0$, with
some contribution in the tail originating from events with FSR. The
background template shape is taken from the single-dissociative events
simulated in \pp collisions with \superchic~v4.0 interfaced with \pythia. These events have a much
wider $\alpha$ distribution than the signal.
The acoplanarity shape is strongly correlated with the transverse momentum of the
$\gamma\gamma$ system, which is driven by the transverse momenta of the initial photons.
For the photons emitted coherently from the nucleus, the transverse momentum is of
order $\hbar c/R_{A} \approx 30$\,\MeV, while typical \pt scale for dissociative
events is of order \GeV. In the case of dilepton production in \pp collisions,
the typical initial \pt scale is about 200~\MeV. The convolution of photon fluxes originating
from either proton or ion with photons emitted from nucleon substructure is always dominated
by the harder spectrum of the latter. Therefore, the shape of the acoplanarity distribution
for dissociative dielectron production in \PbPb collisions can be described by the simulation
of this process in \pp collisions.
The fit to the data is performed in the same intervals of \mee and \Yee
as in the study of the fractions of events in the 0n0n, Xn0n and XnXn categories.
In each bin, a binned maximum-likelihood fitting
procedure is performed separately in three ZDC categories. The
normalisation of the relative background contribution, \fbkg, is taken to be a free
parameter of the fit. The signal fraction is thus $(1-\fbkg)$. For the inclusive sample,
\fbkg is a weighted sum of the results for the 0n0n, Xn0n, and XnXn
categories. The \fbkg fraction accounts for contributions from
dissociative production and exclusive $\tau^+\tau^-$ production, \ggtautau. The latter may contribute to the electron
background, especially when both $\tau$-leptons decay in the electron
channel. The $\tau^+\tau^-$ contribution is estimated using a dedicated MC
sample from \starlight. The resulting background fraction of \ggtautau  events in
the full data sample amounts to 0.1\%. It is found that the shape of
the $\alpha$ distribution for the exclusive $\tau^+\tau^-$ events is similar to
the $\alpha$ distribution for the pure dissociative component. However,
the origin of this shape in $\tau^+\tau^-$ events is due to the presence of the neutrino in $\tau$-lepton decay.
The dissociative contribution, \fdiss, is therefore determined as the background fraction
obtained from the fitting procedure, then reduced by the $\tau^+\tau^-$ background
fraction.
 
\begin{figure}[!htb]
\begin{center}
\includegraphics[width=0.72\textwidth]{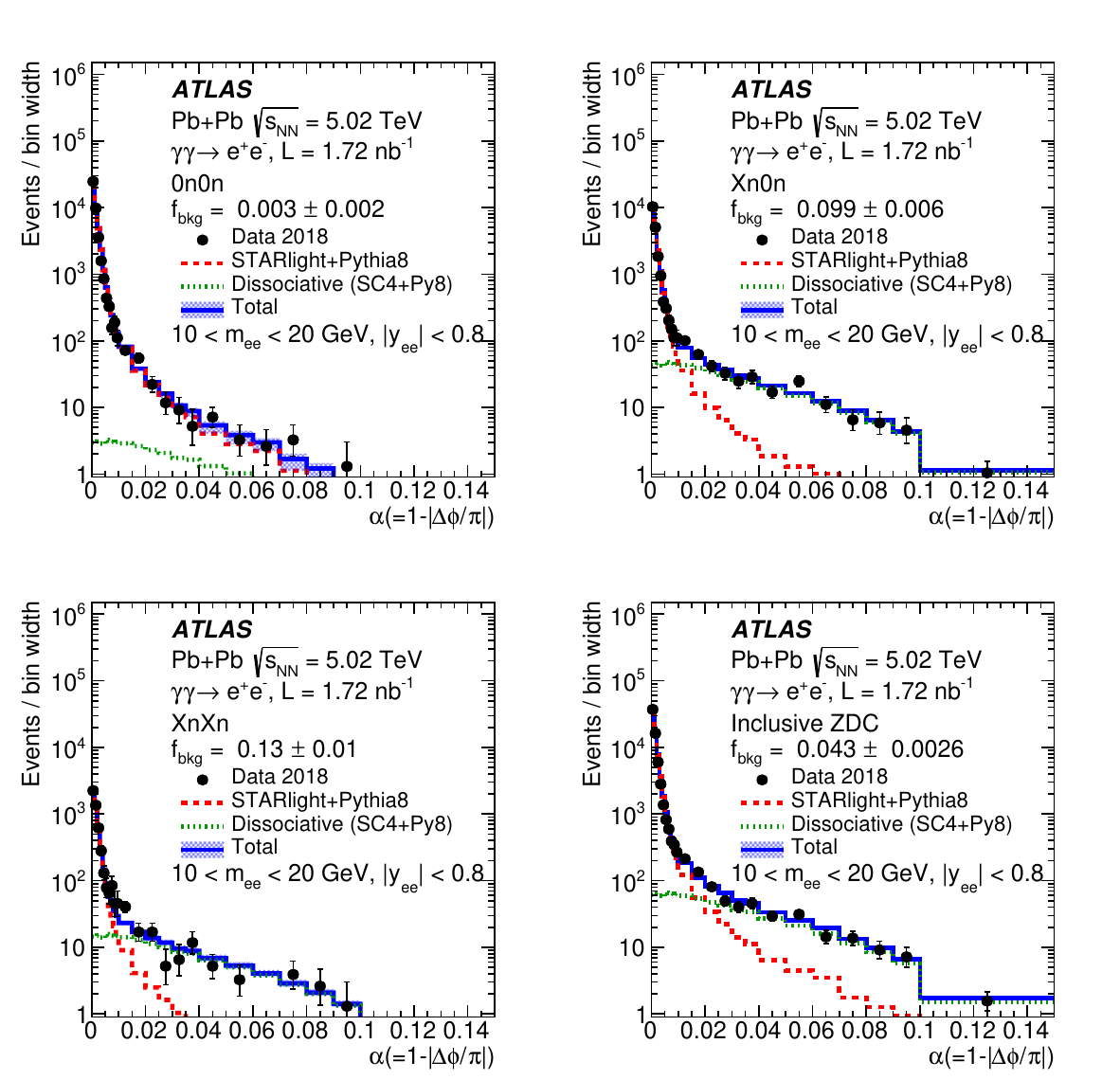}
\caption{Acoplanarity distribution in the
data sample~(markers) of \ggee candidates
selected  with  $10<\mee<20$\,\GeV\ and
$\Yee<0.8$ requirements. The sample is
split into 0n0n (top left), Xn0n (top
right), XnXn (bottom left) and inclusive
(bottom right) categories. The fitted dissociative
background in each category is shown with the green
dashed line, while the prediction for the
signal process is shown by the red
line. The sum of the two components is
shown with the solid blue line. The resulting
estimate of the background fraction in
the data, \fbkg, is given in the legend. The
shaded area represents the total
uncertainty of the sum of signal and background components.}
\label{fig:fitResult0n0n}
\end{center}
\end{figure}

The results of the fitting procedure for the data from the
$10<\mee<20$\,\GeV\ and $\Yee<0.8$ interval  are presented in
Figure~\ref{fig:fitResult0n0n} for three ZDC categories as well as for
the inclusive sample. The \fbkg fraction amounts to $(0.3\pm 0.2)\%$, $
(9.9\pm0.6)\%$, $(13\pm 1)\%$ and $(4.3\pm0.3)\%$ for the 0n0n, Xn0n, XnXn categories,
and the inclusive sample, respectively and increases with \mee and \Yee.

The contribution from $\Upsilon$-meson production is estimated using the
dedicated \starlightPythia samples. Three $\Upsilon$ states,
$\Upsilon(1S), \Upsilon(2S)$, and $\Upsilon(3S)$ are considered. This
background is significant only for \mee below 14\,\GeV\ and amounts to
2.4\% of events satisfying the selection criteria for the \ggee process. The
$\alpha$ distribution of the dielectron candidates from $\Upsilon$
decays is peaked at zero, similarly to the signal shape. It does not
contribute to \fbkg because the resulting fraction is not sensitive to a 2.4\%
change in the acoplanarity peak. Hence, the $\Upsilon$ contribution is subtracted
separately from the data.
 
A background contribution originating from photonuclear processes
occurring in ultraperipheral heavy-ion collisions is largely suppressed by
the trigger requirement limiting the maximum \et deposited in the FCal to 3~\GeV\ per side.
The validity of this assumption was tested by examining the multiplicity distribution of Pixel-tracks.
The fraction of events
that have more than two Pixel-tracks is 1.3\% and consistent with
simulations of \ggee events. This also confirms that the pileup contribution from additional
low-\pt \ggee interactions, which should be below 1\%, can be neglected in the selected data sample.
 
\begin{figure}[!t]
\begin{center}
\includegraphics[width=0.39\textwidth]{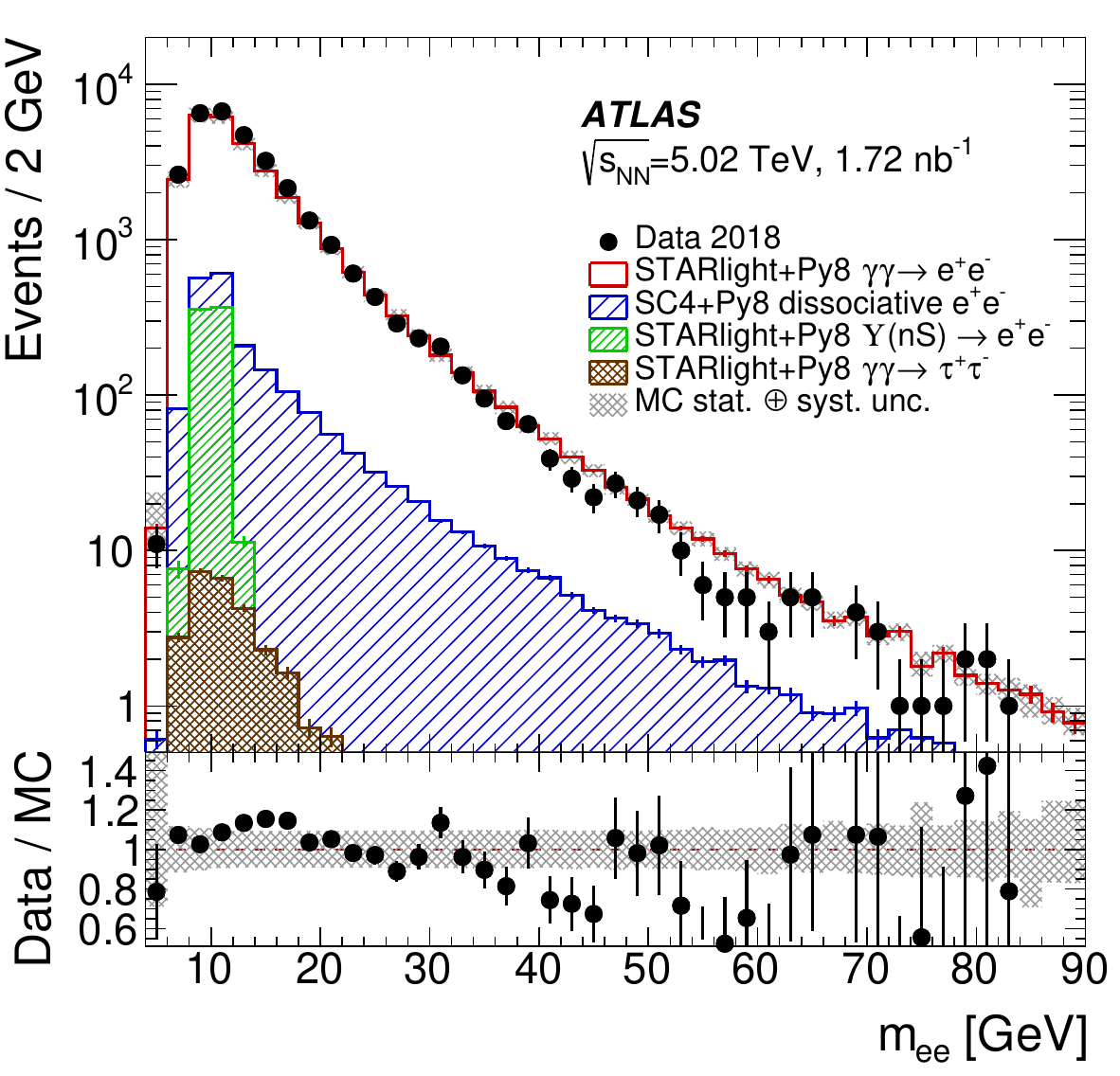}
\includegraphics[width=0.39\textwidth]{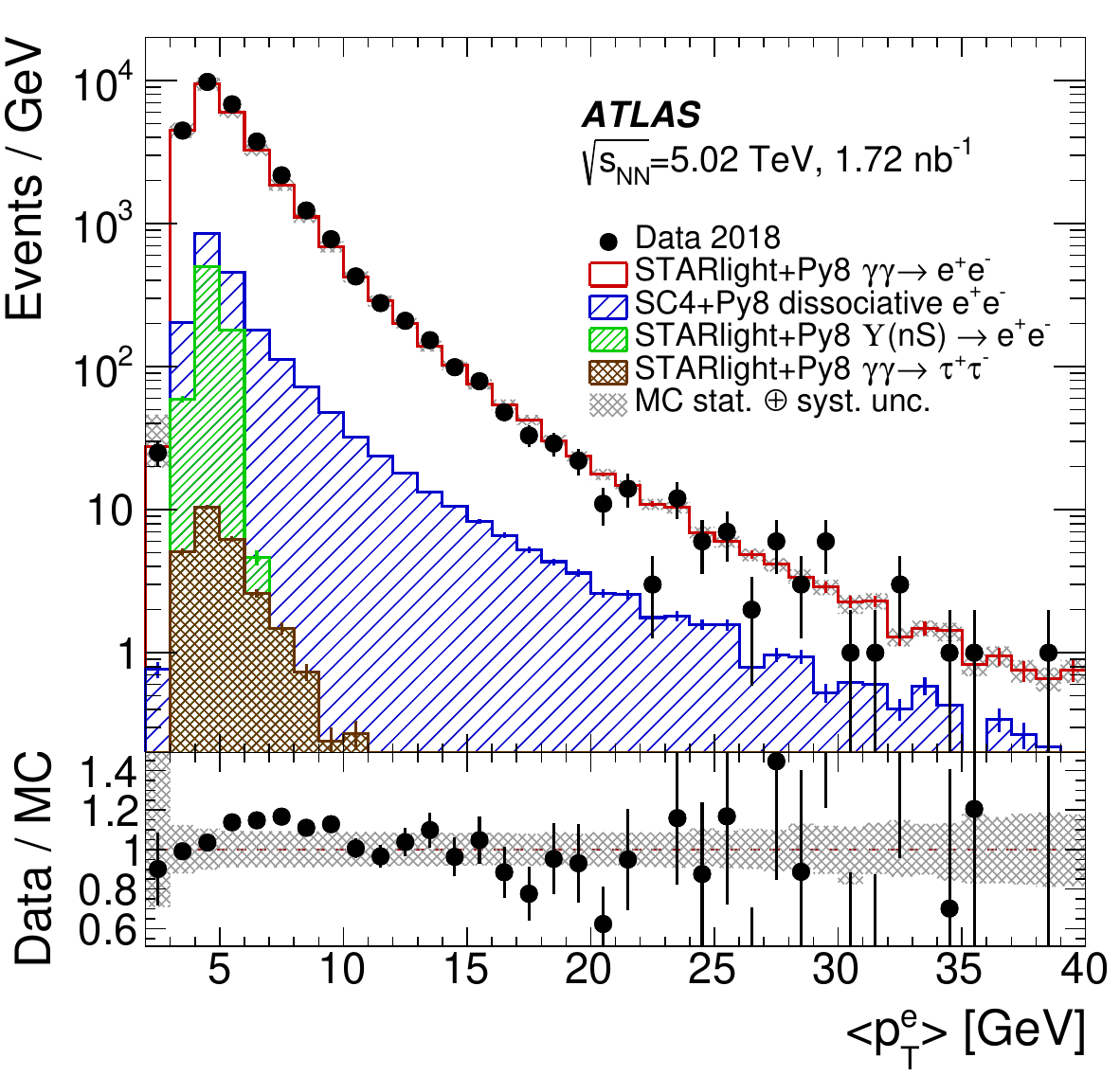}
\includegraphics[width=0.39\textwidth]{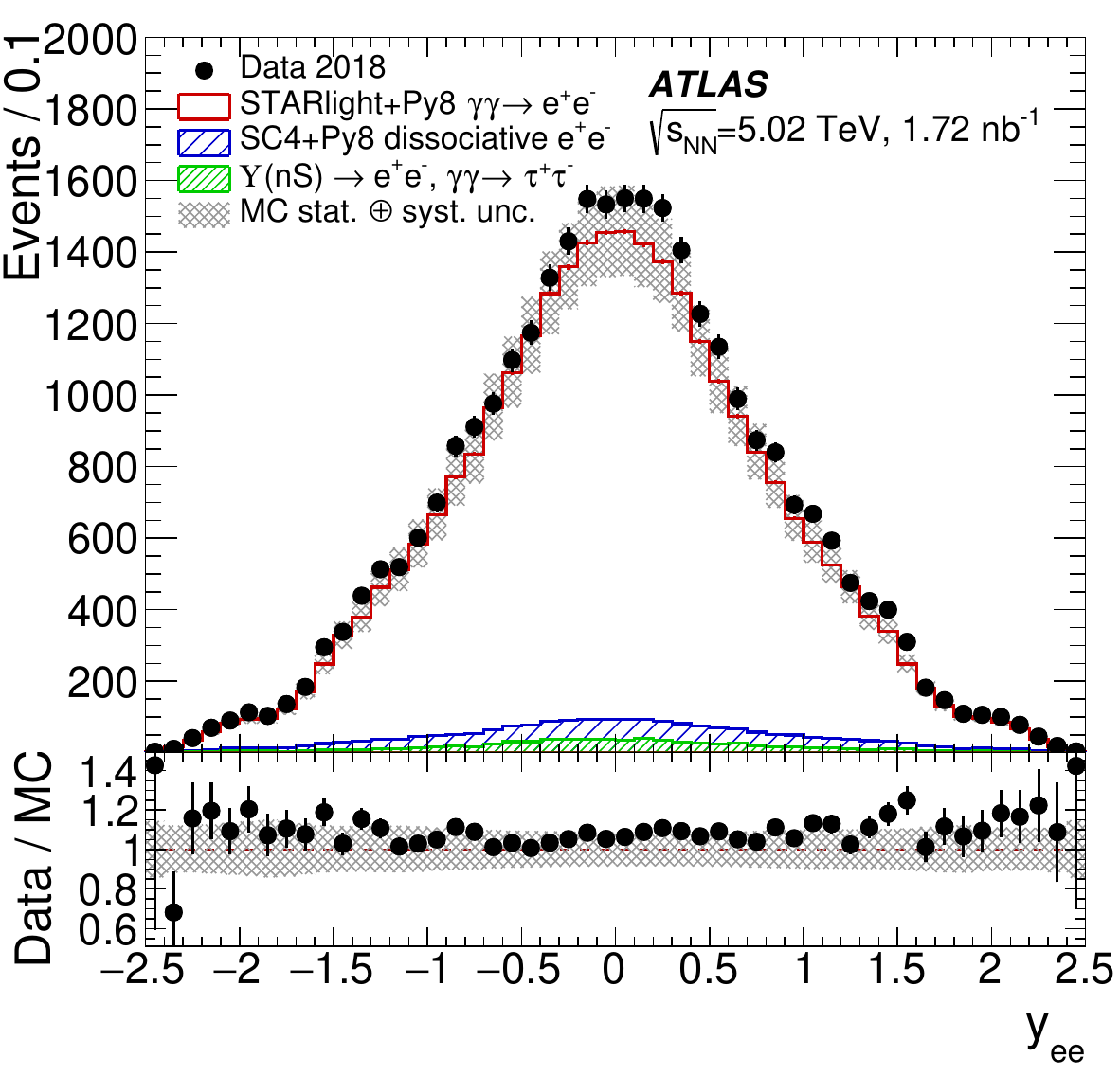}
\includegraphics[width=0.39\textwidth]{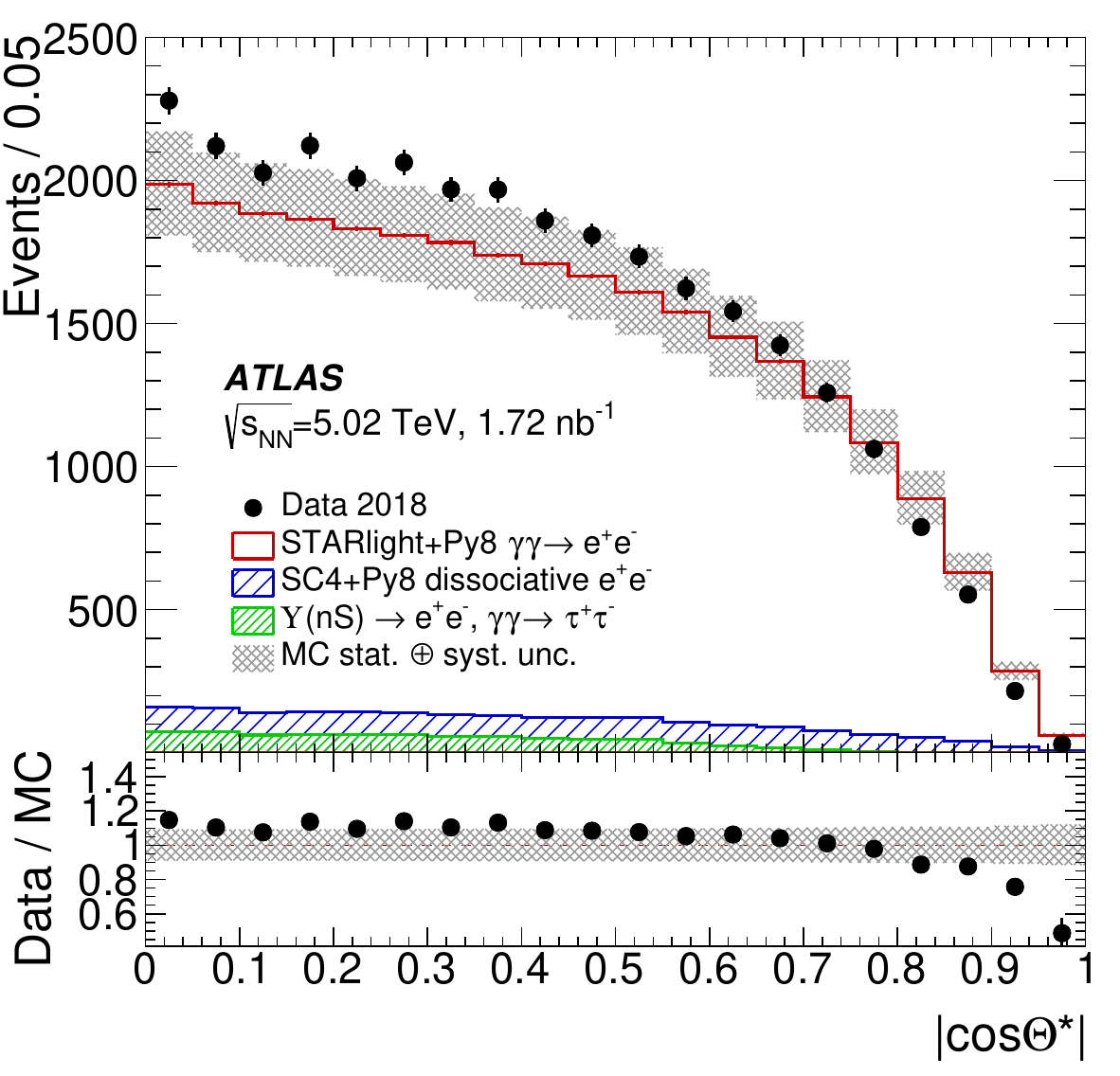}
\includegraphics[width=0.39\textwidth]{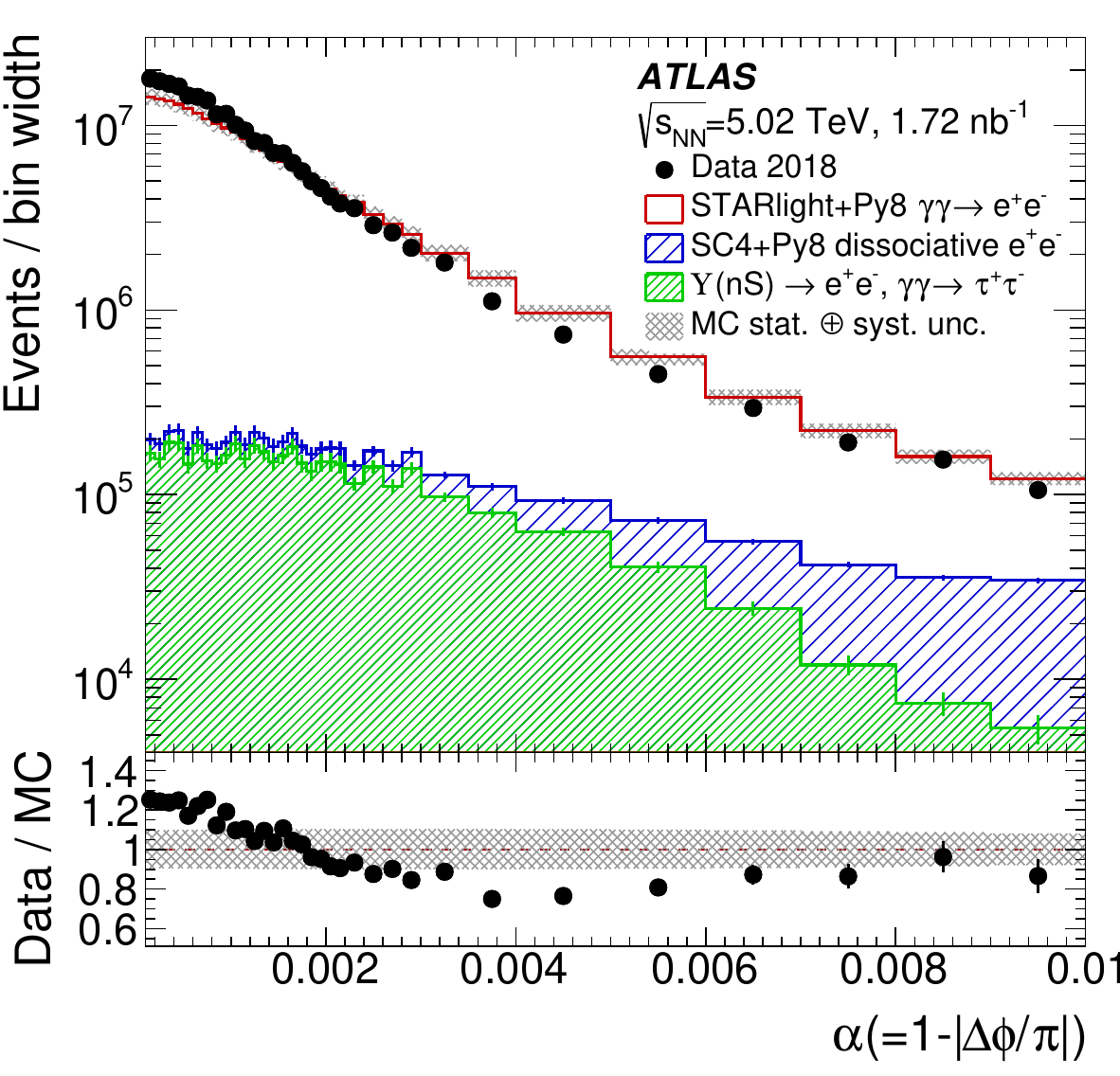}
\caption{Distributions of \mee~(top left),
$\langle \pTe \rangle$~(top right),
$y_{ee}$~(middle left), \costs~(middle
right), and $\alpha$~(bottom) for the
inclusive sample in data and the
MC predictions for signal and background
processes. The lower panels show the ratio of data to
MC simulation. Error bars represent
statistical uncertainties. The
shaded area represents the overall
uncertainty of the total MC prediction. In the $y_{ee}$,
\costs, and $\alpha$ distributions, the $\Upsilon$ and $\tau^+\tau^-$
contributions are shown together. The dissociative contribution
is scaled to constitute the \fdiss fraction from the data fit.  }
\label{fig:controlPlots}
\end{center}
\end{figure}

Figure~\ref{fig:controlPlots} shows control distributions for
reconstructed and uncorrected values of \mee,
$\langle \pTe \rangle$, $y_{ee}$, $\costs= |\tanh(\Delta \eta_{ee}/2)|$, and $\alpha$ for the
inclusive data sample compared with MC predictions including the signal and
background processes.  The trigger decision is not simulated in the MC events.
Instead, the distributions are weighted event-by-event by the parameterised trigger
efficiency and by electron scale factors. The signal, $\Upsilon$ decay and $\tau^+\tau^-$ MC prediction is normalised
to the integrated luminosity in data. The dissociative contribution is scaled to constitute
the \fdiss fraction (determined for the inclusive sample) of the data.
In general, good agreement between data and the sum of the predictions
for signal and background processes is found. On average, the observed
discrepancies are at the level of 10\%--15\% with some exceptions which
are discussed further.
In the \mee distribution the data excess is more strongly
pronounced for \mee between 10 and 20\,\GeV, where the difference
between data and MC simulation is 10\%--20\%.  The data-to-MC ratio drops
below unity for masses above 40\,\GeV. The same features are observed in
the \apTe distribution, with the largest deviations from unity in the
range 5--10\,\GeV. In the $y_{ee}$ distribution, the data excess is smaller, up to 10\%,
in the range from $-1.2$ to 1.2, with larger discrepancies
for higher \Yee values. The data-to-MC ratio in the \costs distribution drops slowly
from 1.2 for $\costs= 0$ to unity at $\costs = 0.75$, and then falls more steeply,
to 0.5 for the largest values of \costs. In the $\alpha$ distribution, a
difference in the overall shape is observed in the full range. This can be explained by a
sensitivity of the results to the \pT spectrum assumed by \starlight, since this spectrum determines
the width of the $\alpha$ distribution. In general, all these
discrepancies tend to be consistent with the observations made
in the ATLAS \ggmumu\ measurement~\cite{ATLAS:2020epq}, where the \starlight
predictions were found to underestimate the measured integrated
fiducial cross-sections by about 10\%.
 
\FloatBarrier
 
\section{Analysis}

The integrated fiducial cross-section for exclusive dielectron production is calculated using the following formula:
\begin{equation*}
\sigma^{\mathrm{fid}} = \frac{N_{\mathrm{data}}-N_{\mathrm{bkg}}}{C \cdot A \cdot L},
\end{equation*}
where:
\begin{itemize}
\item $N_{\mathrm{data}}$ and $N_{\mathrm{bkg}}$ refer to the
number of events in data after event selection and the expected number of
background events in this selected sample,
respectively;
\item $C$ is a correction factor accounting for detector
inefficiencies (including the trigger), calculated as
$N^{\mathrm{fid,cut}}_{\mathrm{MC,reco}}/N^{\mathrm{fid,cut}}_{\mathrm{MC,gen}}$
where $N^{\mathrm{fid,cut}}_{\mathrm{MC,gen}}$ is the number of
generated events passing fiducial requirements of the analysis, while
$N^{\mathrm{fid,cut}}_{\mathrm{MC,reco}}$ is the number of
simulated signal events that also pass the reconstruction-level selection;
\item $A$ is the acceptance correction, used to correct the
result for the exclusion of the calorimeter transition region and extrapolation
from $\etae<2.47$ to $\etae<2.5$; and is calculated as
$N^{\mathrm{fid,cut}}_{\mathrm{MC,gen}}/N^{\mathrm{fid}}_{\mathrm{MC,gen}}$,
where $N^{\mathrm{fid}}_{\mathrm{MC,gen}}$ is the number of
generated events passing all fiducial requirements of the analysis,
except the requirement to exclude the calorimeter transition region;
\item $L$ is the total integrated luminosity.
\end{itemize}
 
Both $N^{\mathrm{fid,cut}}_{\mathrm{MC,gen}}$ and $N^{\mathrm{fid}}_{\mathrm{MC,gen}}$ are
extracted with respect to the generator-level electrons before FSR.
The fiducial region is defined by the following requirements:
$\pTe > 2.5$\,\GeV, $\etae<2.5$, $\mee > 5$\,\GeV, and $\pTee < 2$\,\GeV.
The number of events passing the fiducial selection is $N_\mathrm{data}=30456$.
The dissociative and $\tau^+\tau^-$ background fraction obtained from the fit amounts to 4.5\%.
The $\Upsilon$ background amounts to 2.4\% of all events satisfying the selection
criteria.
 
The selected data sample is corrected in a few subsequent steps in
order to compare it with the theoretical predictions.
In the first step the backgrounds are subtracted. Distributions in the data are reweighted
event-by-event by the factor $(1-\fbkg)$ where \fbkg is the fraction
of background (inclusive in ZDC) from the fit in a given \mee and \Yee
range. For masses above 40\,\GeV, the fraction obtained from the fit in
the full $\Yee<2.4$ range is taken. For events with \mee below 40\,\GeV, the
fraction as a function of \Yee is used.  If \Yee exceeds 2.4, the
fraction from the $1.6<\Yee<2.4$ bin is applied. In the next step, the
background expected from $\Upsilon(nS)$ decays is subtracted.
 
For differential cross-section measurement,
the data are corrected with fiducial correction factors defined
as the fraction of events in each bin which fall into the fiducial
region at generator level. These factors
correct for the events that are reconstructed within the
fiducial region, but fall outside it at the generator level. They are parameterised
using the reconstructed kinematic variables. They deviate from unity
at the subpercent level, so their impact is marginal.
After this, the reconstructed data are unfolded using a
Bayesian-inspired iterative procedure~\cite{DAGOSTINI1995487} with one iteration for all
distributions implemented in the RooUnfold
package~\cite{Adye:2011gm}, using response matrices derived
from signal MC samples.  The number of iterations is chosen to minimise
the resulting statistical uncertainty and at the same time provide good closure.
A closure test based on the signal MC samples is performed to validate
the unfolding procedure. The signal sample is split into two parts.
The first part is used to fill the response matrices, while the second one
is unfolded. The ratio of the unfolded yields to the generated yields deviates from unity
by 1\% at most.
 
Finally, the distributions are divided by the
luminosity as well as the product of correction factors, $C \times A$, which account for
detector inefficiencies as well as acceptance losses. They are
determined for each bin of the unfolded distribution as the fraction
of events that pass the fiducial requirements at reconstruction level, in events that
pass them at generator level. The $C \times A$ factors are parameterised using generator-level
kinematics, but then weighted by trigger and reconstruction efficiency scale factors,
evaluated at reconstruction level. The average $C$ and $A$ factors
amount to 0.087 and 0.878, respectively.
 
\section{Systematic uncertainties}

The following systematic uncertainties are considered in the cross-section measurement.
The total scale factors for the electron reconstruction and identification
efficiency~\cite{ATLAS:2019qmc} are varied upwards and downwards coherently over
the full kinematic range, as a conservative estimate.
The data-driven trigger efficiency, which is the product of the L1
efficiency, the Pixel-veto efficiency, and the forward transverse
energy requirement efficiency, is increased and decreased by its total
uncertainty.
To assess known uncertainties in the EM energy scale and
energy resolution, the calibrations are varied by factors
determined in 13\,\TeV\ \pp collisions~\cite{ATLAS:2019qmc}.
The background contributions are increased and decreased
by their total uncertainties. The dissociative
backgrounds are dominated by their statistical uncertainties from the fit.
The systematic uncertainties are also evaluated, and the largest contribution is
related to the shape of the signal template. This is estimated
using data from the 0n0n category as a signal distribution for the Xn0n and XnXn
categories. The background template shape uncertainty is estimated by adding
the double dissociative component. The uncertainty in the expected $\Upsilon$ yields
is dominated by both the efficiency scale factors and the
EM energy scale. Given the small contribution from $\Upsilon$ production,
the theoretical uncertainties of its cross-section are considered to have
negligible impact on the final measurement.
 
For the differential cross-sections, additional systematic uncertainties are
related to the unfolding procedures.
The MC sample is split in two, with one subsample used to determine the response
matrix and the other treated as a simulated data set.  The differences between the
generated and unfolded yields are treated as a systematic uncertainty.
Similarly, the sensitivity to the Bayesian prior is tested by reweighting the simulated
data set to agree with the reconstructed data. Again, the differences between simulated and
reconstructed yields in this closure test are applied as an uncertainty.
While the primary unfolding is evaluated in one dimension rather than two,
a cross-check is performed using the response in two dimensions. For each
of the unfolded variables, every other one is used in the second dimension,
but with the number of bins reduced to four~(three) for \mee and \apTe (\Yee and \costs)
to compensate for the limited number of events in the MC sample. The three resulting
two-dimensional cross-sections are projected to one dimension and compared with
the nominal results for each unfolded variable. The largest variations in each bin
are included as an uncertainty. Finally, the spectra are evaluated in the 0n0n category,
using the fractions determined in Section~\ref{sec:analysis}, but evaluated for
generator-level $m_{ee}$ and $|y_{ee}|$ (as opposed to the
reconstructed values). The differences are found to be within 1\%--2\%.
 
The uncertainty in the integrated luminosity of the data sample is
2.0\%. It is derived from the calibration of the luminosity
scale using $x$--$y$ beam-separation scans, following a methodology
similar to that detailed in Ref.~\cite{DAPR-2013-01}, and using the
LUCID-2 detector for the baseline luminosity
measurements \cite{LUCID2}.
 
\begin{figure}[!htb]
\begin{center}
\includegraphics[width=0.48\textwidth]{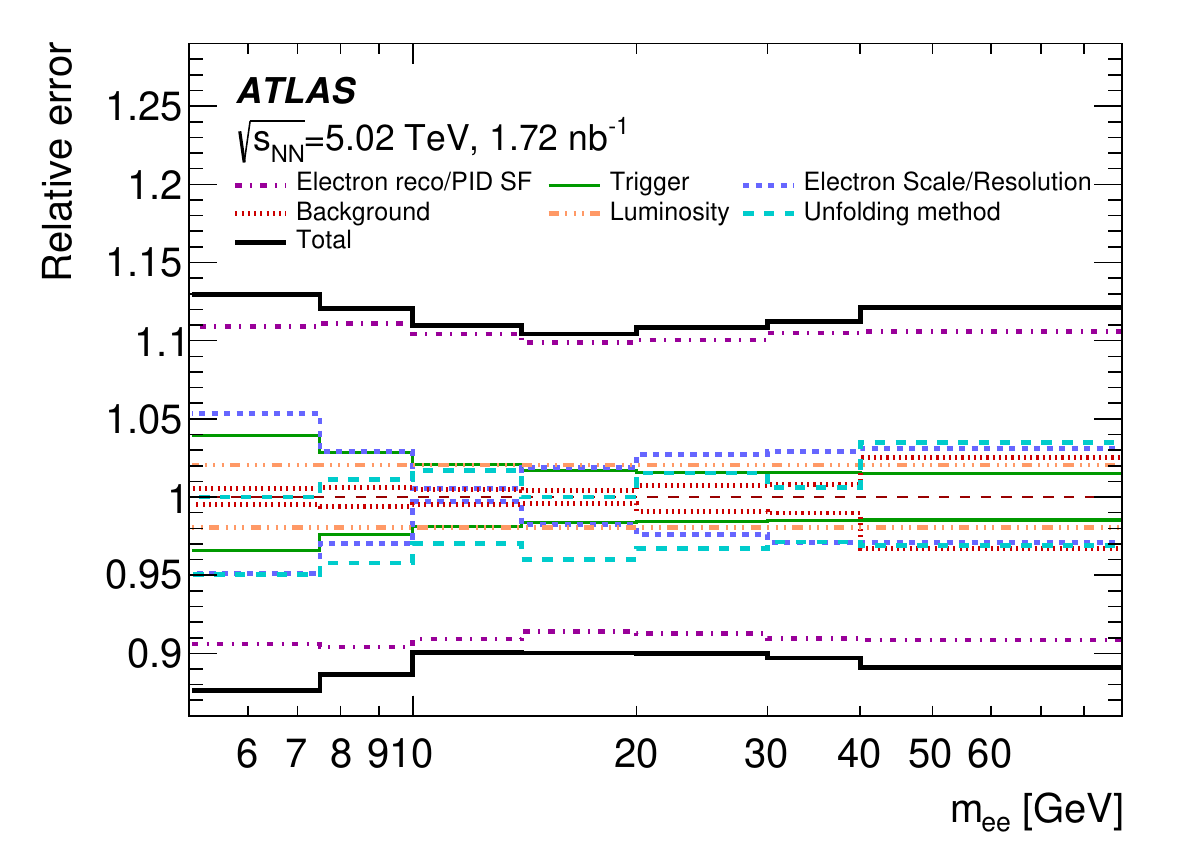}
\includegraphics[width=0.48\textwidth]{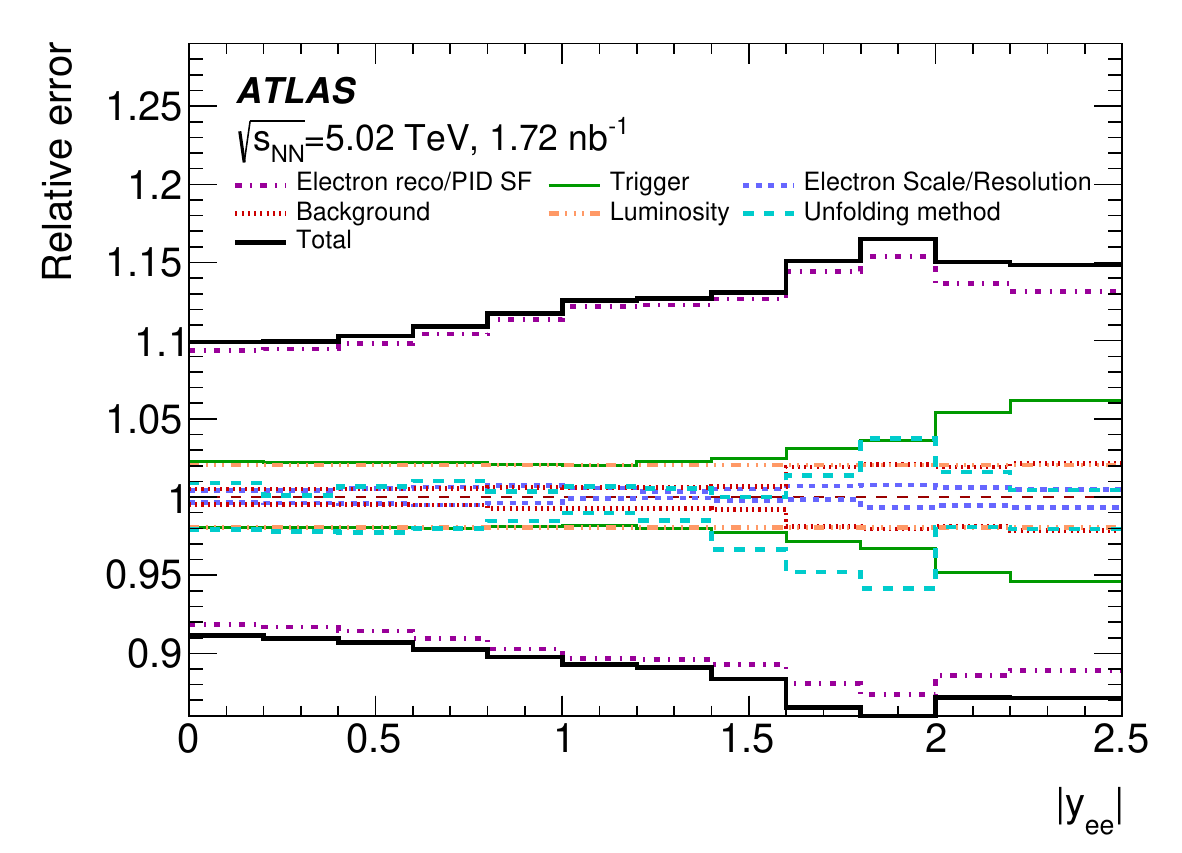}
\caption{Breakdown of relative systematic uncertainties in the differential
cross-section as a function of \mee~(left) and \Yee~(right).}
\label{fig:syst1}
\end{center}
\end{figure}

A summary of the systematic uncertainties as a function of \mee and \Yee is shown in
Figure~\ref{fig:syst1}. The dominant source of uncertainty arises from the uncertainties in the
electron scale factors. They are at the level of 10\%--11\% in the whole range of \mee,
and rise from 9\% at \Yee $\approx$ 0 to about 15\% for \Yee close to 2.
The systematic uncertainty from the trigger efficiency is approximately 2\%
for \mee above 10\,\GeV\ and \Yee below 1.6. It rises to 4\% for smaller \mee,
and to 6\% for the highest \Yee values. The uncertainty related to the energy scale and resolution is below
1\% in the whole range of \Yee, but exceeds this value in some \mee bins, reaching
approximately 5\% for the lowest \mee values.
The background uncertainties are within 1\%--3\% and increase slightly with increasing \mee and \Yee.
Uncertainties related to unfolding procedures do not exhibit such clear dependencies in \mee
and \Yee. They are mostly within the 2\%--3\% range but exceed this value, up to 5\%, at
intermediate \mee and \Yee.
 
\section{Results}
\label{sec:result}
 
The total integrated fiducial cross-section is measured to be $215 \pm 1\text{(stat.)}~\substack{+23\\-20}\text{(syst.)} \pm 4\text{(lumi.)}\,\mu$b.
The \starlight prediction for the total integrated fiducial cross-section is 196.9\,$\mu$b, while the \superchic prediction
is 235.1\,$\mu$b. Both predictions are statistically compatible with the measurement.
 
\begin{figure}[!htb]
\begin{center}
\includegraphics[width=0.48\textwidth]{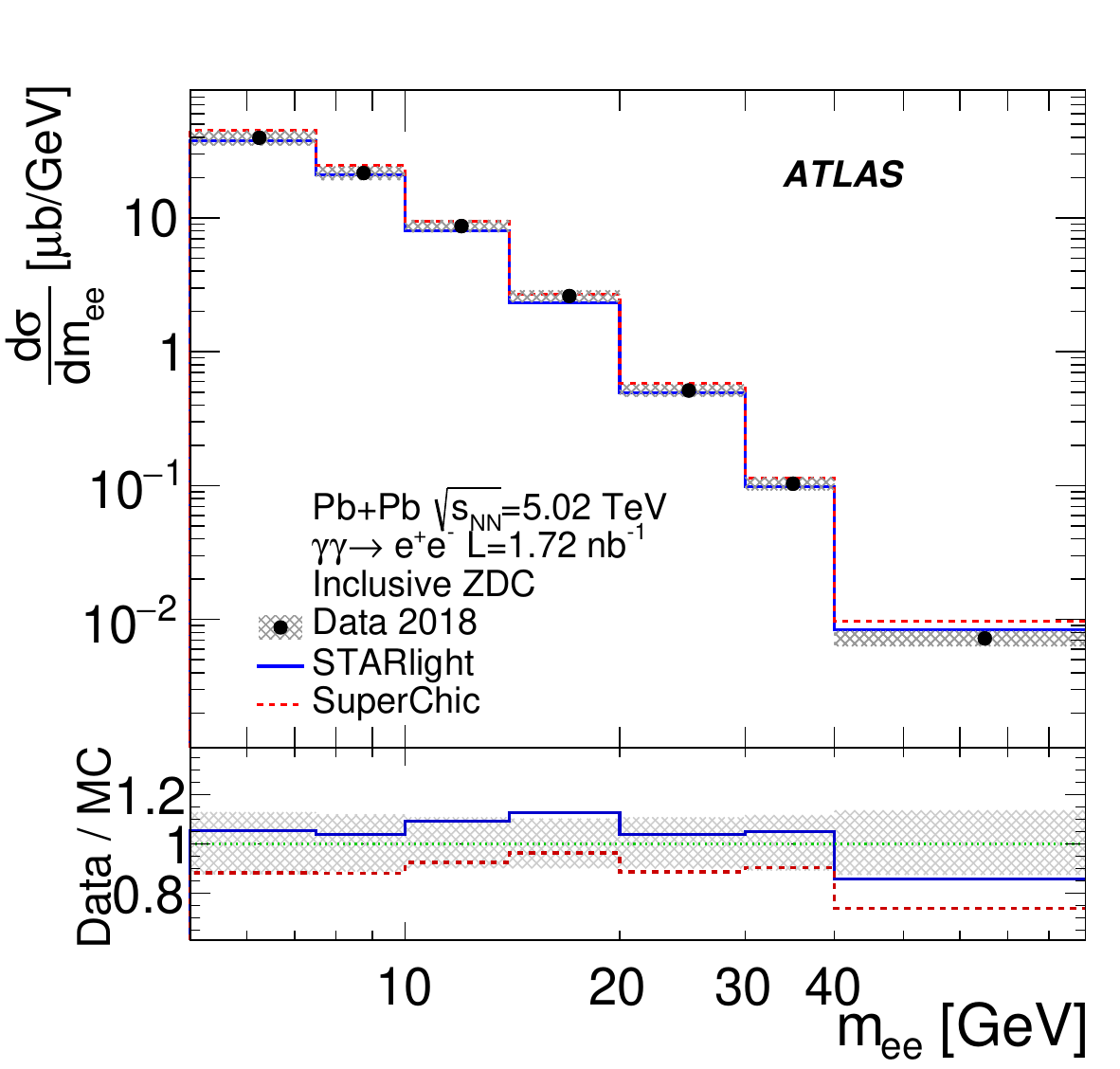}
\includegraphics[width=0.48\textwidth]{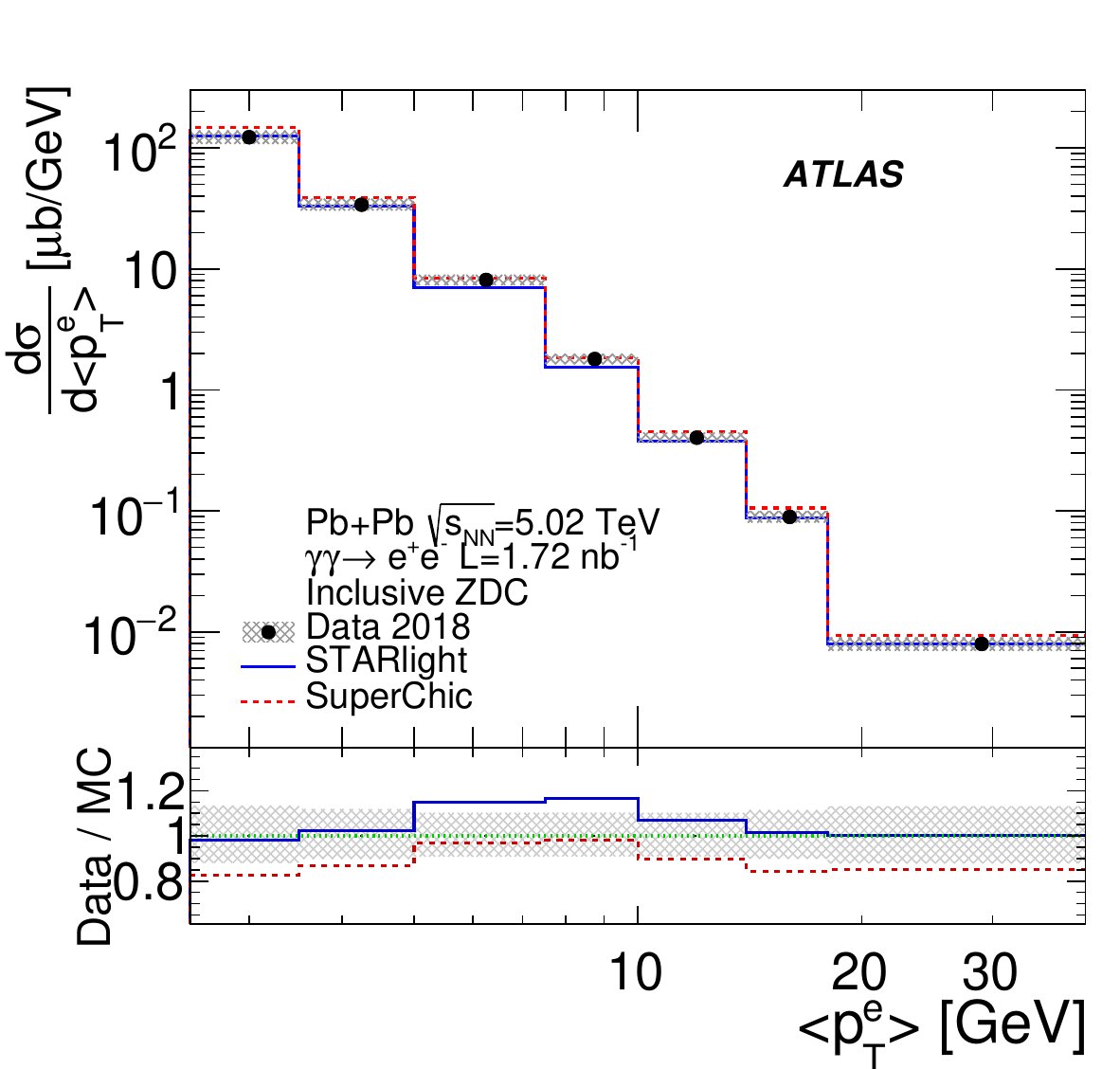}\\
\includegraphics[width=0.48\textwidth]{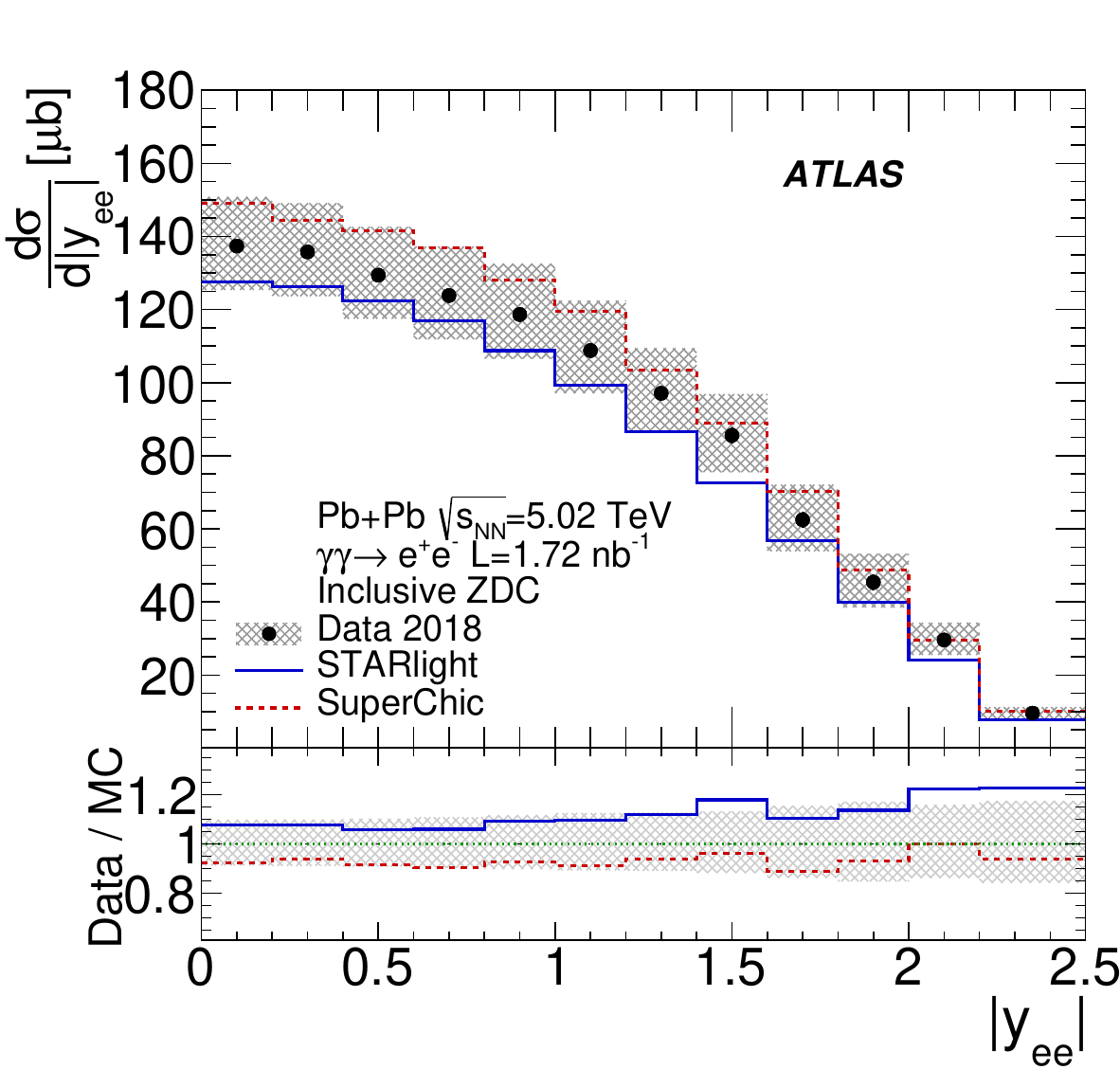}
\includegraphics[width=0.48\textwidth]{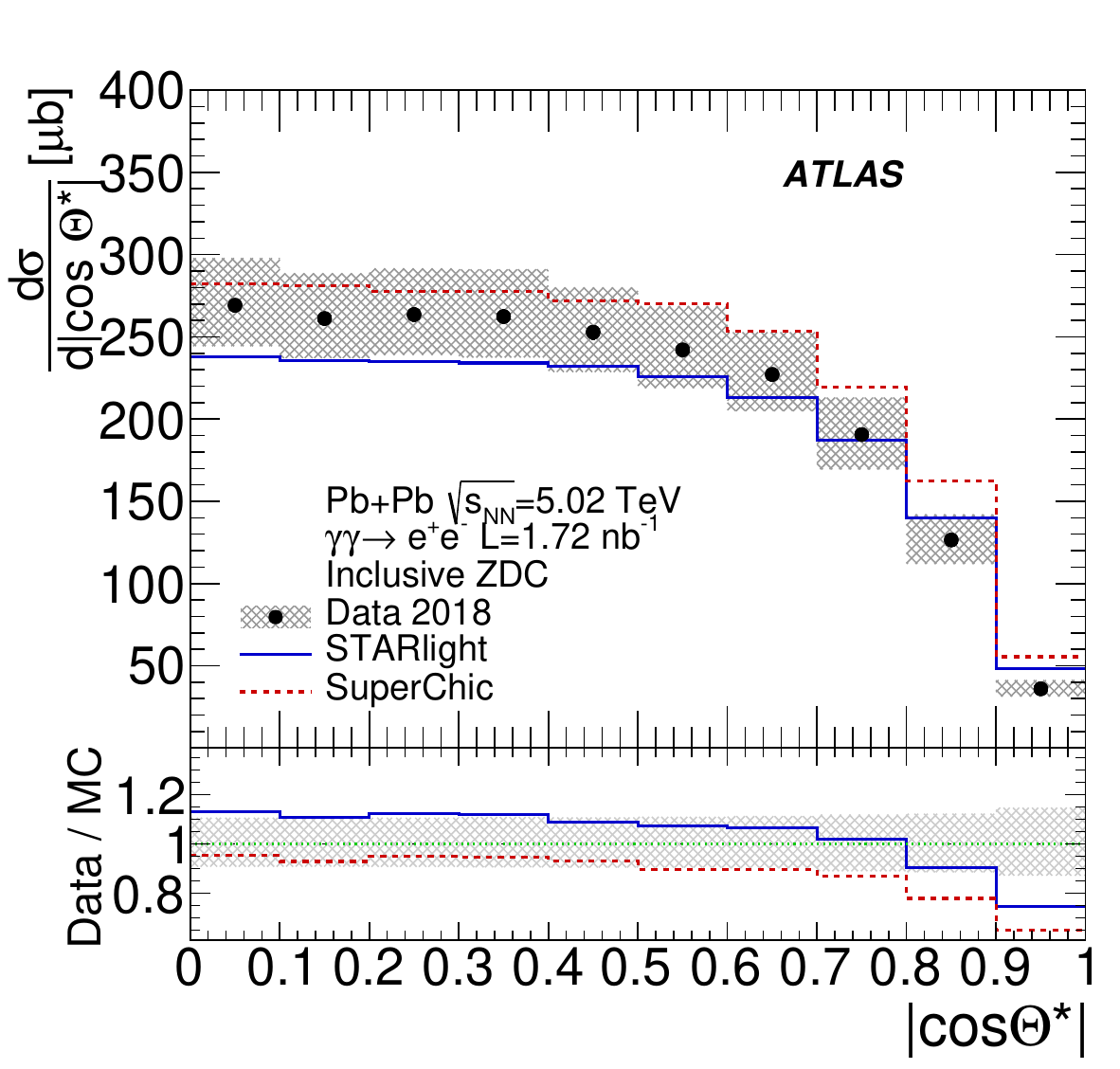}
\caption{Fully corrected differential cross-sections measured inclusively in ZDC categories
for exclusive dielectron production,
$\gamma\gamma\rightarrow e^+e^-$, as a function of
\mee, \apTe, \Yee and \costs for data~(dots) and MC
predictions from \starlight~(solid blue) and
\superchic~(dashed red). Bottom panels
present the ratios of data to MC predictions. The
shaded area represents the total
uncertainty of the data, excluding the 2\% luminosity uncertainty.}
\label{fig:xsecResult}
\end{center}
\end{figure}

The differential cross-sections for exclusive dielectron production
are presented as a function of \mee, \apTe, \Yee, and \costs in
Figure~\ref{fig:xsecResult}. The cross-sections are measured inclusively
in the ZDC categories. The results are corrected for detector
inefficiency and resolution effects, and are compared with \starlight~v3.13 and
\superchic~v3.05 predictions for the signal \ggee process.
The bottom panel in each plot shows the ratio of the unfolded data to MC
predictions. On average the \starlight predictions underestimate
the data by about 10\%--15\%, while \superchic predictions are higher by about the same
amount. The \starlight and \superchic predictions tend to have very similar shapes. The
difference in the absolute normalisation of the two predictions is
due to different approaches in the calculation of the initial photon flux.
The predictions describe the shape of the
data well, except at high \Yee and high \costs. The differences are
more pronounced for \mee between 10 and 20\,\GeV, and for \apTe between 5 and
10\,\GeV. The ratio of data to \starlight rises from about 1.1 to 1.2 as \Yee increases
from 0 to 2.5. For \costs close to 0, the data-to-\starlight ratio reaches its
largest value, around 1.15, and then slowly decreases to about 1.05 for
$\costs= 0.8$. The ratio falls more steeply in the last two bins
of \costs and drops below unity to 0.75 and 0.65 for \starlight and
\superchic respectively.
The measured cross-section in the highest \costs bin is 1.8~(2.7) standard deviations
below the theory prediction from \starlight (\superchic).
There is a plausible proposal that higher-order scattering processes (involving more
than two photons in the initial state) are relevant and would tend to reduce the
predicted cross-sections by the observed discrepancies~\cite{Zha:2021jhf}.
 
\begin{figure}[!htb]
\begin{center}
\includegraphics[width=0.48\textwidth]{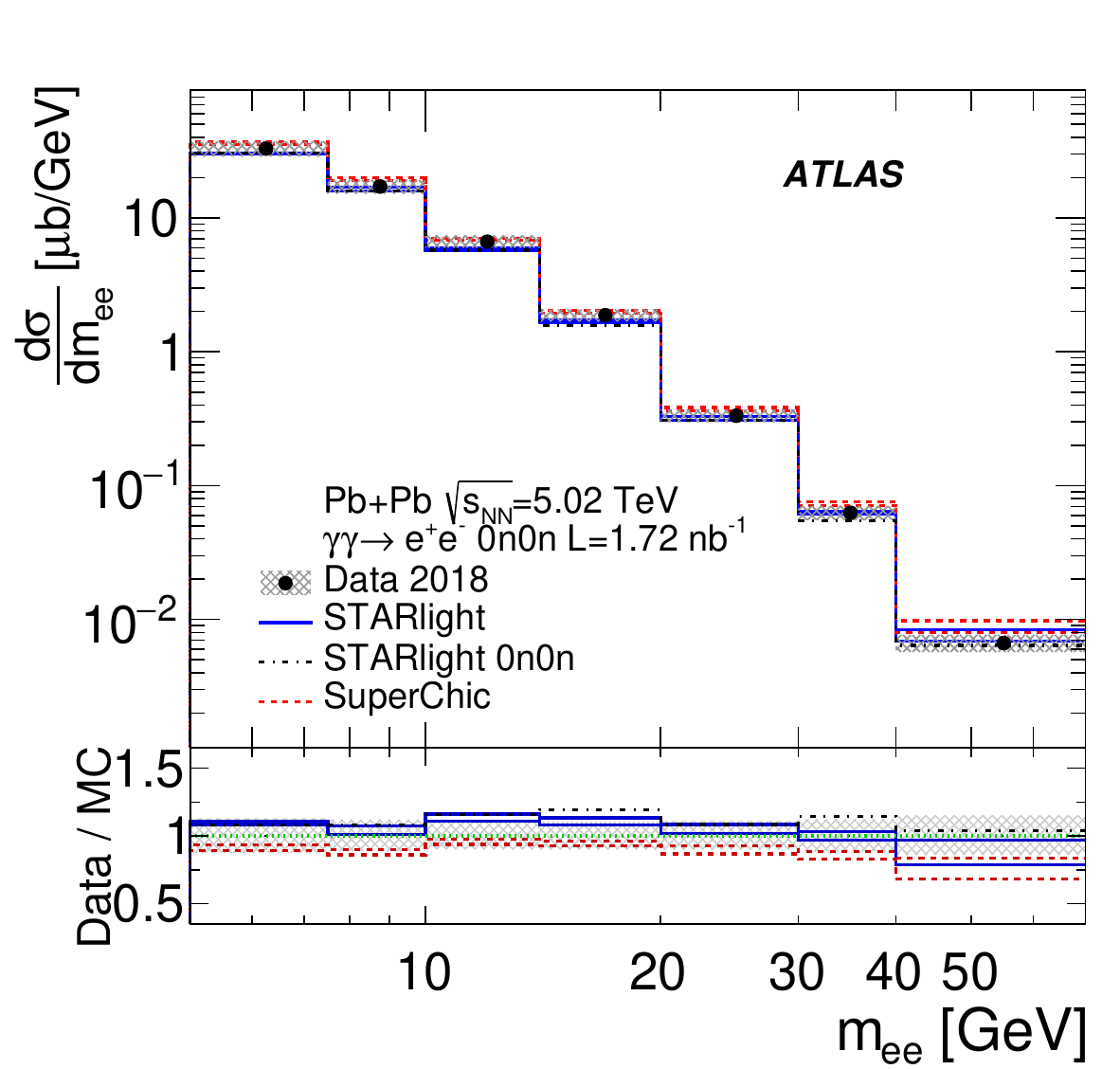}
\includegraphics[width=0.48\textwidth]{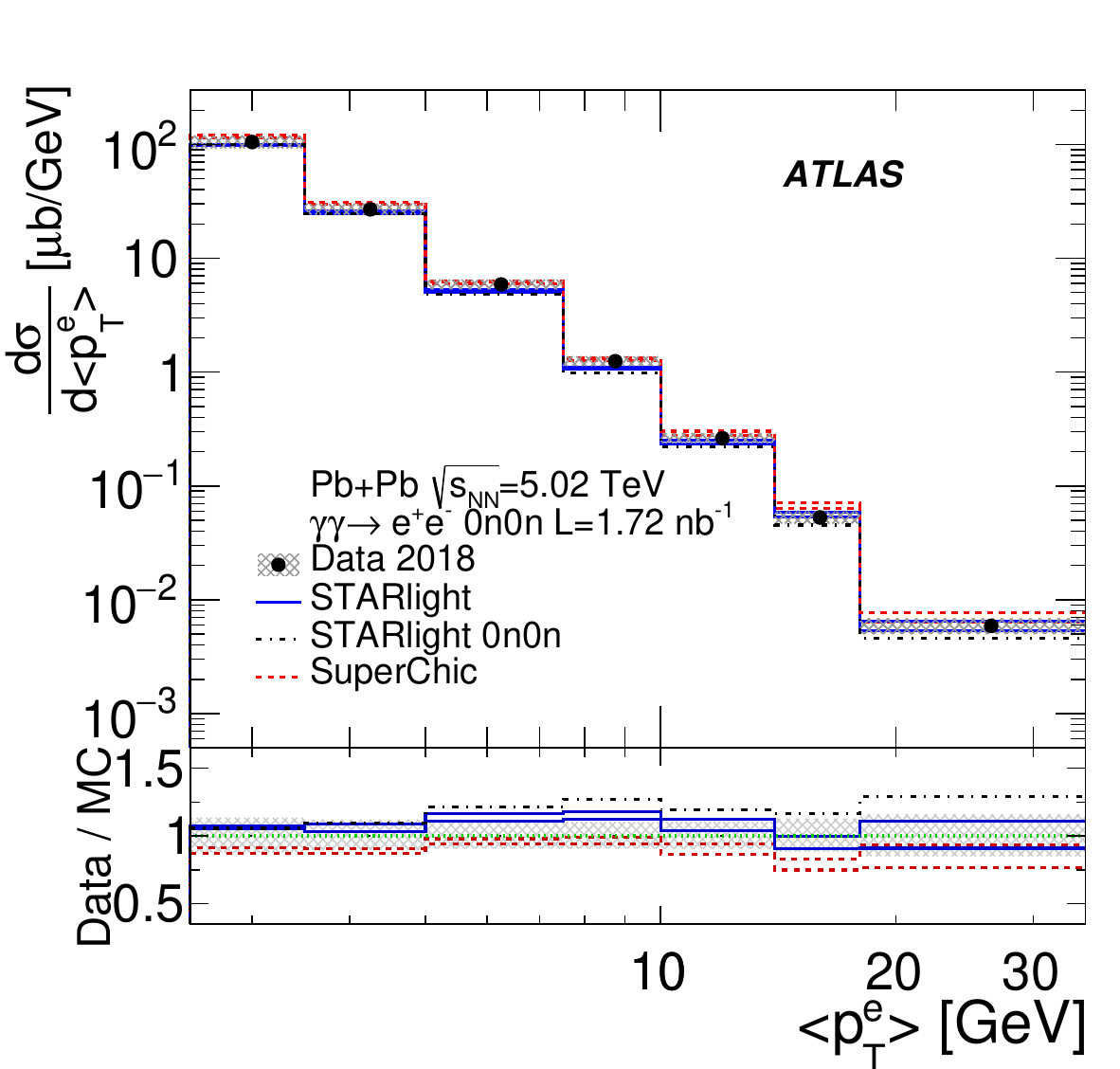}\\
\includegraphics[width=0.48\textwidth]{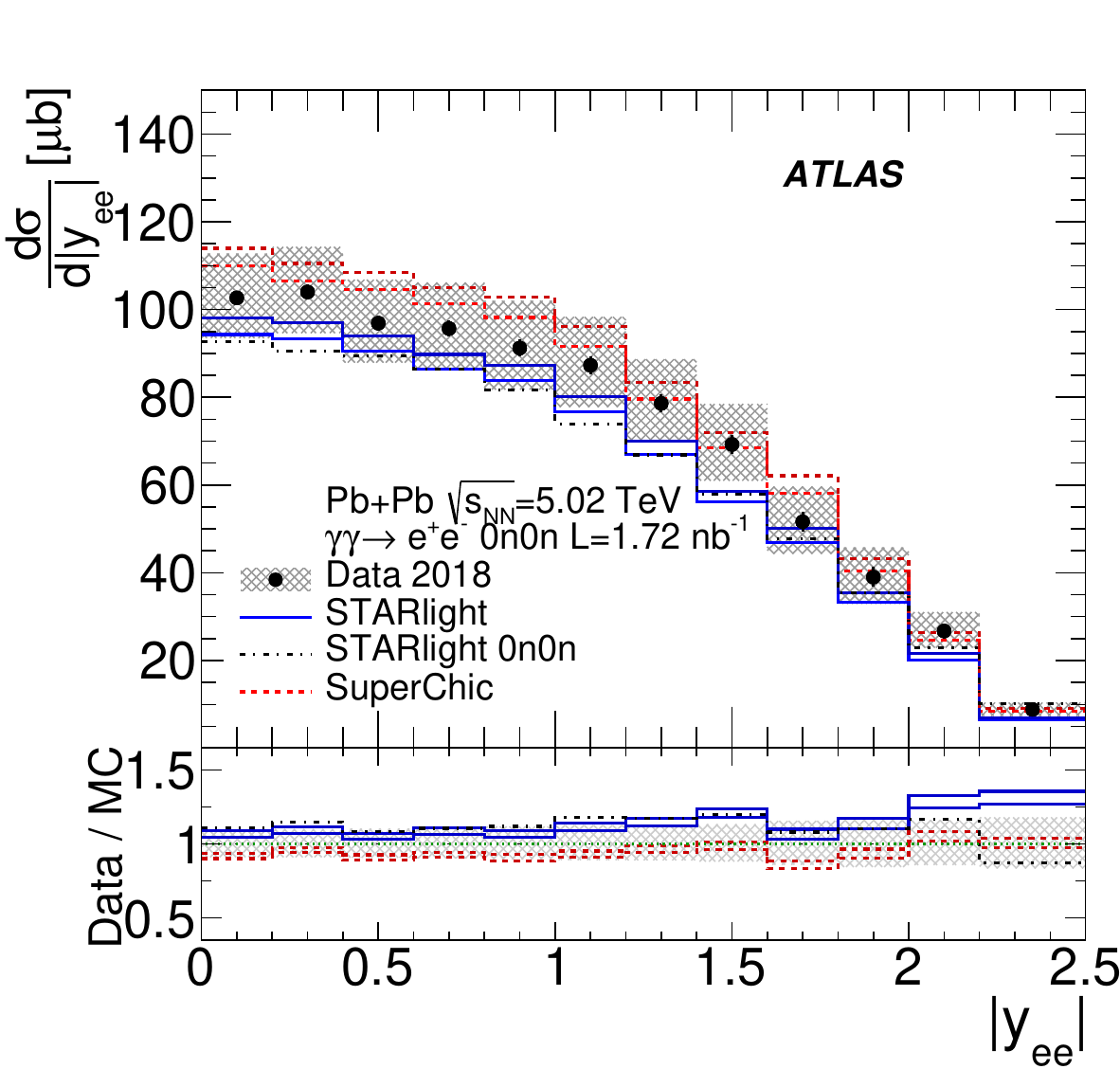}
\includegraphics[width=0.48\textwidth]{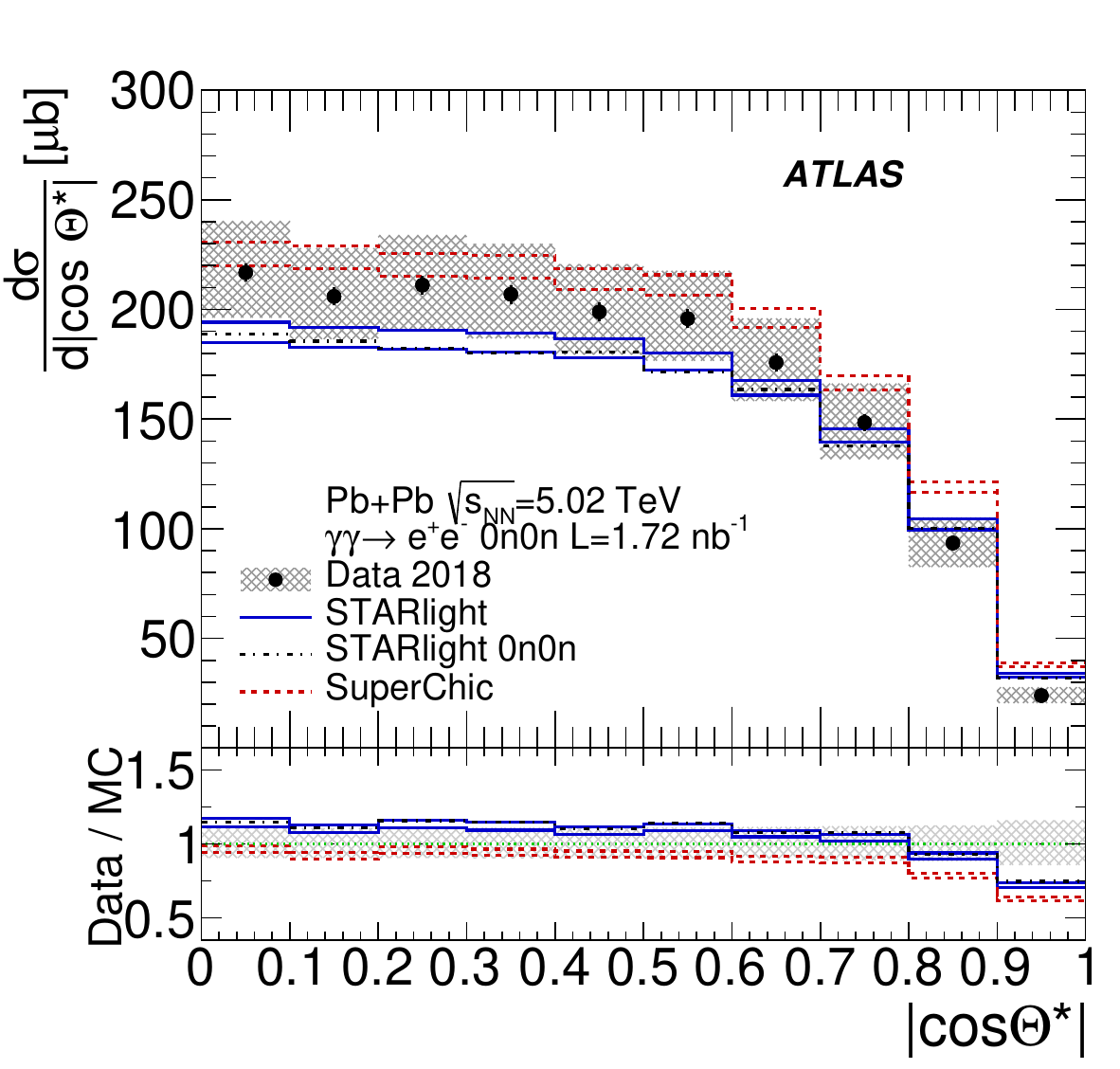}
\caption{Fully corrected differential
cross-sections measured for the 0n0n category for exclusive dielectron
production as a function of \mee, \apTe,
\Yee and \costs. The cross-sections are
compared with MC predictions from \starlight (solid blue)
and \superchic~v3.05 (dashed red), each represented
by two lines reflecting systematic
variations. Also, a dedicated prediction
from \starlight for the 0n0n category
(dashed-dotted black) is
shown. The bottom panels show the ratios of data to
predictions. The shaded area
represents the total uncertainty of the
data, excluding the 2\% luminosity uncertainty.}
\label{fig:xsecResult0n0n}
\end{center}
\end{figure}

The differential cross-sections  as a function of \mee, \apTe, \Yee
and \costs for the 0n0n category are presented in
Figure~\ref{fig:xsecResult0n0n}. They are compared with the MC
predictions from \starlight~v3.13 and \superchic~v3.05. Both simulated
samples were produced inclusively and reweighted to the 0n0n category
using the measured fractions presented in Fig.~\ref{fig:f0n0n}.
The reweighting was performed event-by-event based on the generator
level values of \mee and \Yee. Each theory
prediction is represented by two curves reflecting the systematic
variations of the measured 0n0n fractions. \starlight can also
generate a prediction conditional on the presence of neutron emission
in one or both directions. These dedicated predictions
from \starlight for the 0n0n category are shown in the same plots. That
prediction agrees well with the shape of the inclusive \starlight prediction
corrected for the measured 0n0n fractions, but is systematically lower
by 2\%--3\% for $\Yee<1.4$. The general conclusions from this comparison between MC
predictions and data are consistent with the inclusive case.
Agreement between data and MC events is generally
better for lower \Yee and \costs values, i.e.\ involving lower-energy initial-state photons.

\FloatBarrier

\section{Conclusions}

A measurement of the cross-section for exclusive dielectron
production, \yyee, is performed using \Lint of ultraperipheral \PbPb
collision data at $\sqn = 5.02$\,\TeV\ recorded by the ATLAS detector at
the LHC. The cross-section is corrected for detector efficiency,
acceptance losses, and background contributions. The backgrounds from
dissociative processes, $\Upsilon$ decays, and $\tau^+\tau^-$ production are subtracted,
with the first contribution estimated using a template fit to the
acoplanarity distribution.  After all corrections, the integrated
cross-section for the \yyee process in the fiducial region, defined
by $\pTe > 2.5$\,\GeV, $\etae < 2.5$, $\mee > 5$\,\GeV, and $\pTee < 2$\,\GeV\
requirements, is measured to be $215 \pm 1\text{(stat.)}~\substack{+23\\-20}\text{(syst.)} \pm 4\text{(lumi.)}\,\mu$b.
Within experimental
uncertainties the data are in good agreement with the QED predictions
from \starlight~v3.13 and \superchic~v3.05. The differential cross-sections
are presented as a function of \mee, \apTe, \Yee and \costs, both with
and without requirements on forward neutron activity.
The differential results are compared with the
predictions from \starlight~v3.13, which are generally lower than measured cross-sections, and from \superchic~v3.05, which tend to be higher.
In general, the shapes of the distributions agree well, but some
systematic differences are observed.
In particular, the discrepancy between
data and the \starlight prediction rises with higher \Yee, similarly to what ATLAS observed
previously in \ggmumu production. For  $\costs\approx 1$, the measured cross-section is 1.8~(2.7) standard deviations
below the theory prediction from \starlight (\superchic).
Agreement between data and MC simulation is generally better for lower \Yee and \costs values.
 
\section*{Acknowledgements}

% The next lines are included from the .//acknowledgements/Acknowledgements.tex input file

We thank CERN for the very successful operation of the LHC, as well as the
support staff from our institutions without whom ATLAS could not be
operated efficiently.
 
We acknowledge the support of
ANPCyT, Argentina;
YerPhI, Armenia;
ARC, Australia;
BMWFW and FWF, Austria;
ANAS, Azerbaijan;
CNPq and FAPESP, Brazil;
NSERC, NRC and CFI, Canada;
CERN;
ANID, Chile;
CAS, MOST and NSFC, China;
Minciencias, Colombia;
MEYS CR, Czech Republic;
DNRF and DNSRC, Denmark;
IN2P3-CNRS and CEA-DRF/IRFU, France;
SRNSFG, Georgia;
BMBF, HGF and MPG, Germany;
GSRI, Greece;
RGC and Hong Kong SAR, China;
ISF and Benoziyo Center, Israel;
INFN, Italy;
MEXT and JSPS, Japan;
CNRST, Morocco;
NWO, Netherlands;
RCN, Norway;
MEiN, Poland;
FCT, Portugal;
MNE/IFA, Romania;
MESTD, Serbia;
MSSR, Slovakia;
ARRS and MIZ\v{S}, Slovenia;
DSI/NRF, South Africa;
MICINN, Spain;
SRC and Wallenberg Foundation, Sweden;
SERI, SNSF and Cantons of Bern and Geneva, Switzerland;
MOST, Taiwan;
TENMAK, T\"urkiye;
STFC, United Kingdom;
DOE and NSF, United States of America.
In addition, individual groups and members have received support from
BCKDF, CANARIE, Compute Canada and CRC, Canada;
PRIMUS 21/SCI/017 and UNCE SCI/013, Czech Republic;
COST, ERC, ERDF, Horizon 2020 and Marie Sk{\l}odowska-Curie Actions, European Union;
Investissements d'Avenir Labex, Investissements d'Avenir Idex and ANR, France;
DFG and AvH Foundation, Germany;
Herakleitos, Thales and Aristeia programmes co-financed by EU-ESF and the Greek NSRF, Greece;
BSF-NSF and MINERVA, Israel;
Norwegian Financial Mechanism 2014-2021, Norway;
NCN and NAWA, Poland;
La Caixa Banking Foundation, CERCA Programme Generalitat de Catalunya and PROMETEO and GenT Programmes Generalitat Valenciana, Spain;
G\"{o}ran Gustafssons Stiftelse, Sweden;
The Royal Society and Leverhulme Trust, United Kingdom.
 
The crucial computing support from all WLCG partners is acknowledged gratefully, in particular from CERN, the ATLAS Tier-1 facilities at TRIUMF (Canada), NDGF (Denmark, Norway, Sweden), CC-IN2P3 (France), KIT/GridKA (Germany), INFN-CNAF (Italy), NL-T1 (Netherlands), PIC (Spain), ASGC (Taiwan), RAL (UK) and BNL (USA), the Tier-2 facilities worldwide and large non-WLCG resource providers. Major contributors of computing resources are listed in Ref.~\cite{ATL-SOFT-PUB-2021-003}.

% End of text imported from the .//acknowledgements/Acknowledgements.tex input file

\clearpage

\printbibliography
 
\clearpage
% ATLAS Collaboration author list
% Reference date of HION-2021-16 is 2022-03-15
% Author list last updated on date 07-JUL-23
% Data extracted on 07-Jul-2023 for paper reference HION-2021-16
% at 2:52pm
 
\begin{flushleft}
\hypersetup{urlcolor=black}
{\Large The ATLAS Collaboration}

\bigskip

\AtlasOrcid[0000-0002-6665-4934]{G.~Aad}$^\textrm{\scriptsize 101}$,
\AtlasOrcid[0000-0002-5888-2734]{B.~Abbott}$^\textrm{\scriptsize 119}$,
\AtlasOrcid[0000-0002-7248-3203]{D.C.~Abbott}$^\textrm{\scriptsize 102}$,
\AtlasOrcid[0000-0002-1002-1652]{K.~Abeling}$^\textrm{\scriptsize 55}$,
\AtlasOrcid[0000-0002-8496-9294]{S.H.~Abidi}$^\textrm{\scriptsize 29}$,
\AtlasOrcid[0000-0002-9987-2292]{A.~Aboulhorma}$^\textrm{\scriptsize 35e}$,
\AtlasOrcid[0000-0001-5329-6640]{H.~Abramowicz}$^\textrm{\scriptsize 150}$,
\AtlasOrcid[0000-0002-1599-2896]{H.~Abreu}$^\textrm{\scriptsize 149}$,
\AtlasOrcid[0000-0003-0403-3697]{Y.~Abulaiti}$^\textrm{\scriptsize 116}$,
\AtlasOrcid[0000-0003-0762-7204]{A.C.~Abusleme~Hoffman}$^\textrm{\scriptsize 136a}$
% ...
\bigskip
\\

% ...
$^{29}$Physics Department, Brookhaven National Laboratory, Upton NY; United States of America.\\
% ...
$^{35}$$^{(a)}$Facult\'e des Sciences Ain Chock, R\'eseau Universitaire de Physique des Hautes Energies - Universit\'e Hassan II, Casablanca;$^{(b)}$Facult\'{e} des Sciences, Universit\'{e} Ibn-Tofail, K\'{e}nitra;$^{(c)}$Facult\'e des Sciences Semlalia, Universit\'e Cadi Ayyad, LPHEA-Marrakech;$^{(d)}$LPMR, Facult\'e des Sciences, Universit\'e Mohamed Premier, Oujda;$^{(e)}$Facult\'e des sciences, Universit\'e Mohammed V, Rabat;$^{(f)}$Institute of Applied Physics, Mohammed VI Polytechnic University, Ben Guerir; Morocco.\\
% ...
$^{55}$II. Physikalisches Institut, Georg-August-Universit\"{a}t G\"ottingen, G\"ottingen; Germany.\\
% ...
$^{101}$CPPM, Aix-Marseille Universit\'e, CNRS/IN2P3, Marseille; France.\\
$^{102}$Department of Physics, University of Massachusetts, Amherst MA; United States of America.\\
% ...
$^{116}$Department of Physics, New York University, New York NY; United States of America.\\
% ...
$^{119}$Homer L. Dodge Department of Physics and Astronomy, University of Oklahoma, Norman OK; United States of America.\\
% ...
$^{136}$$^{(a)}$Departamento de F\'isica, Pontificia Universidad Cat\'olica de Chile, Santiago;$^{(b)}$Millennium Institute for Subatomic physics at high energy frontier (SAPHIR), Santiago;$^{(c)}$Instituto de Investigaci\'on Multidisciplinario en Ciencia y Tecnolog\'ia, y Departamento de F\'isica, Universidad de La Serena;$^{(d)}$Universidad Andres Bello, Department of Physics, Santiago;$^{(e)}$Instituto de Alta Investigaci\'on, Universidad de Tarapac\'a, Arica;$^{(f)}$Departamento de F\'isica, Universidad T\'ecnica Federico Santa Mar\'ia, Valpara\'iso; Chile.\\
% ...
$^{149}$Department of Physics, Technion, Israel Institute of Technology, Haifa; Israel.\\
$^{150}$Raymond and Beverly Sackler School of Physics and Astronomy, Tel Aviv University, Tel Aviv; Israel.\\
% ...
$^{a}$ Also Affiliated with an institute covered by a cooperation agreement with CERN.\\
$^{b}$ Also at An-Najah National University, Nablus; Palestine.\\
$^{c}$ Also at Borough of Manhattan Community College, City University of New York, New York NY; United States of America.\\
$^{d}$ Also at Bruno Kessler Foundation, Trento; Italy.\\
$^{e}$ Also at Center for High Energy Physics, Peking University; China.\\
$^{f}$ Also at Centro Studi e Ricerche Enrico Fermi; Italy.\\
% ...

\end{flushleft}

% Created with Glance <Atlas.Glance@cern.ch>

\end{document}